\documentclass[fleqn,usenatbib]{mnras}

\usepackage{newtxtext,newtxmath}

\usepackage[T1]{fontenc}
\usepackage{ae,aecompl}

\usepackage{graphicx}	
\usepackage{amsmath}	
\usepackage{amssymb}	
\usepackage{xcolor}

\title[Multiple Populations in Hodge 11 and NGC 2210]{Exploring the nature and synchronicity of early cluster formation in the Large Magellanic Cloud IV: Evidence for Multiple Populations in Hodge 11 and NGC 2210}

\author[Gilligan et al.]{
Christina K. Gilligan,$^{1}$\thanks{E-mail: christina.k.gilligan.gr@dartmouth.edu}
Brian Chaboyer,$^{1}$
Jeffrey D. Cummings,$^{2}$
Dougal Mackey,$^{3}$w
\newauthor
Roger E. Cohen,$^{4}$
Doug Geisler,$^{5}$ $^{6}$ $^{7}$
Aaron J. Grocholski,$^{8}$
M. C. Parisi,$^{9}$ 
\newauthor
Ata Sarajedini,$^{10}$
Paolo Ventura,$^{11}$
Sandro Villanova,$^{5}$
Soung-Chul Yang,$^{12}$
\newauthor
and Rachel Wagner-Kaiser$^{13}$
\\
$^{1}$Department of Physics and Astronomy, Dartmouth College, Hanover, NH 03784, USA\\
$^{2}$Center for Astrophysical Sciences, Johns Hopkins University, 3400 N. Charles Street, Baltimore, MD 21218, USA\\
$^{3}$Research School of Astronomy \& Astrophysics, Australian National University, Canberra, ACT 2611, Australia\\
$^{4}$Space Telescope Science Institute, Baltimore, MD 21218, USA \\
$^{5}$Departamento de Astronom\' ia, Casilla 160-C ,Universidad de Concepci\' on, Chile \\
$^{6}$Instituto de Investigaci\' on Multidisciplinario en Ciencia y Tecnolog\' ia, Universidad de La Serena, Avenida Ra\' ul Bitr\' an S/N, La Serena, Chile \\
$^{7}$Departamento de F\' isica y Astronom\' ia, Facultad de Ciencias, Universidad de La Serena, Av. Juan Cisternas 1200, La Serena, Chile \\ 
$^{8}$Department of Physics, American University, Washington, DC 20016, USA \\
$^{9}$Observatorio Astron\' omico, Universidad Nacional de C\' ordoba, Laprida 854, C\' ordoba, CP 5000, Argentina \\
$^{10}$Florida Atlantic University, 777 Glades Rd, SE-43, Room 256, Boca Raton, FL 33431, USA \\
$^{11}$INAF-Osservatorio Astronomico di Roma, Via Frascati 33, 00078 Monteporzio Catone (Roma), Italy \\ 
$^{12}$Korea Astronomy and Space Science Institute (KASI), Daejeon 305-348, Korea \\
$^{13}$Department of Astronomy, University of Florida, 211 Bryant Space Science Center, Gainesville, FL 32611, USA
}

\date{Accepted XXX. Received YYY; in original form ZZZ}

\pubyear{2015}

\begin{document}
\label{firstpage}
\pagerange{\pageref{firstpage}--\pageref{lastpage}}
\maketitle

\begin{abstract}
We present a multiple population search in two old Large Magellanic Cloud (LMC) Globular Clusters, Hodge 11 and NGC 2210. This work uses data from the Advanced Camera for Surveys and Wide Field Camera 3 on the Hubble Space Telescope from programme GO-14164 in Cycle 23. Both of these clusters exhibit a broadened main sequence with the second population representing ($20 \pm \! \sim \! 5$)\% for NGC 2210 and ($30 \pm \! \sim \! 5$)\% for Hodge 11. In both clusters, the smaller population is redder than the primary population, suggesting CNO variations. Hodge 11 also displays a bluer second population in the horizontal branch, which is evidence for helium enhancement. However, even though NGC 2210 shows similarities to Hodge 11 in the main sequence, there does not appear to be a second population on NGC 2210's horizontal branch. This is the first photometric evidence that ancient LMC Globular Clusters exhibit multiple stellar populations. 
\end{abstract}




\section{Introduction} \label{sec:intro}
Globular clusters (GCs) are among the oldest objects in the Universe. GCs are assumed to be major building blocks of galaxies, with many pre-dating their host and forming at very low metallicities.  In order to relate GC properties to the host galaxy's history, we first need to understand the formation of GCs.  As our understanding of GCs has become more sophisticated, however, it has lead to more questions about their properties and formation, as we find GCs are more complex than previously believed.

It had been thought that GCs were single, Simple Stellar Populations and therefore were able to be described by a single isochrone. This has been proven to not be the case. Nearly every Galactic GC studied photometrically or spectroscopically exhibits some degree of multiple populations \citep{Bastian}. The key features of these multiple populations are: the populations are close in age, there are anti-correlations between O-Na, C-N, e.g., but rarely differences in Fe abundances \citep{Carretta}, and the populations appear to be discrete. The Advanced Camera for Surveys (ACS) Globular Cluster Survey project \citep{Sarajedini} allowed a very detailed and homogeneous study of many Galactic GCs that had never been achieved before. Ages and distances were measured to a less than 10\% precision as done in \citet{Marin}, \citet{ACSWagner-Kaiser}, \citet{Vandenberg}, \citet{Dotter}, and others. One discovery in the project was the evidence for multiple populations in the  Galactic GC NGC 1851 \citep{Milone2008}.

The formation mechanism for multiple populations in GCs is still difficult to explain and account for the vast variety of observational constraints. Proposed creation mechanisms include fast rotating massive stars, massive interacting binaries, supermassive stars, and AGB stars. All of these formation scenarios suffer serious drawbacks. Interested readers are directed to \citet{Renzini} and \citet{Bastian} for a in-depth discussion of the successes and failures of each formation mechanism. One of the main issues with fast rotating massive stars and massive interacting binaries is that they do not produce discrete multiple populations. The AGB pollution mechanism's main drawback is getting the AGB pollution abundances to match the abundances found in the other populations. None of the current scenarios are able to fully explain the observations.

Split main sequences (MS) have been found in many massive Milky Way (MW) GCs \citep[e.g.][]{Piotto,Anderson,NGC6397}. There have also been recent studies looking at young, massive LMC GCs where researchers found evidence for photometric multiple populations in these GCs \citep[e.g.][]{Milone2016,Milone2017}. A few studies have been done for GCs in the SMC including \citet{Paper1}, \citet{Paper2}, and \citet{Paper3} which examine five SMC GCs of varying ages. Through photometry, the authors find that four out of the five GCs exhibit multiple populations. The clusters with multiple populations were either old or intermediate age ($\sim$ 6 Gyr). The one SMC cluster examined that did not have evidence for multiple populations is around 1.5 Gyr. 

In \citet{Martocchia}, six LMC GCs with ages ranging from about 1.5-11 Gyr are analysed. All of the clusters older than 2 Gyr exhibit multiple populations, while clusters younger than this typically do not. The apparent strong dependence of multiple populations on age implies that there is some evolutionary effect causing the multiple populations, along with environmental effects. 

Some old LMC GCs have been studied spectroscopically, including Hodge 11 and NGC 2210 which are discussed here.  \citet{Mateluna} analysed four stars in Hodge 11 and found a [Na/Fe] range between -0.37 and 0.03 dex. NGC 2210 has been previously studied in \citet{Mucciarelli2010} and \citet{Mucciarelli2009}. These two studies along with a third study \citep{Mucciarelli2008} showed that NGC 2210 and the other LMC GCs examined are similar to Galactic GCs in that they have Na-O anticorrelations, which is primary evidence for the presence of multiple populations in these clusters. The authors also found that similarly to most MW GCs, old LMCs are $\alpha$-enhanced.

However, until this work, there have been no photometric studies of the oldest LMC GCs designed to detect multiple populations, and so it is not clear if multiple populations are a ubiquitous feature of GC formation, or something unique to the MW. This work examines ancient LMC GCs in a similar manner as the ACS Globular Cluster Survey, with homogeneous photometry of six LMC GCs with the Hubble Space Telescope (programme GO-14164 in Cycle 23).

\section{Observations and Data Reduction}

There are 15 known ancient ($\sim$ 13 Gyr or older) GCs in the LMC \citep{OldLMC}. The six LMC GCs examined in this project are far from the bar of the LMC, with the two clusters analysed in this paper (NGC 2210 and Hodge 11) being $\sim$ 4 $\deg$ from the bar. For the interested reader, Figure 1 of \citet{Rachel} shows the spatial configuration of these LMC GCs. We avoided clusters superimposed on the LMC bar due to the effects of crowding becoming too large to obtain high-quality photometry of faint main sequence stars, which is required for studies of multiple populations. 

\begin{figure*}
  \centering
  \begin{tabular}{cc}
    \includegraphics[width=\columnwidth]{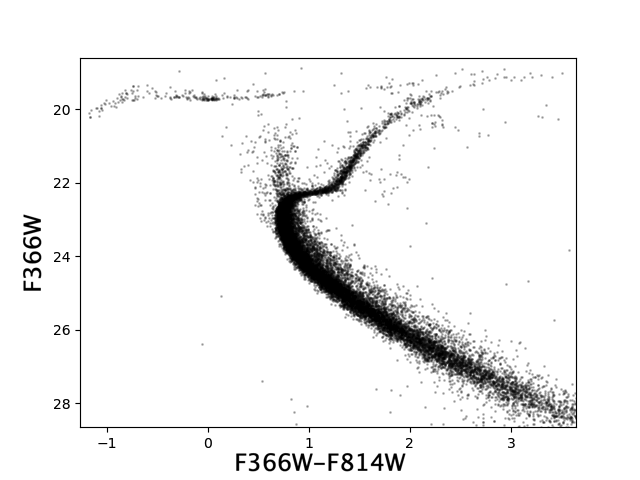} 
        \includegraphics[width=\columnwidth]{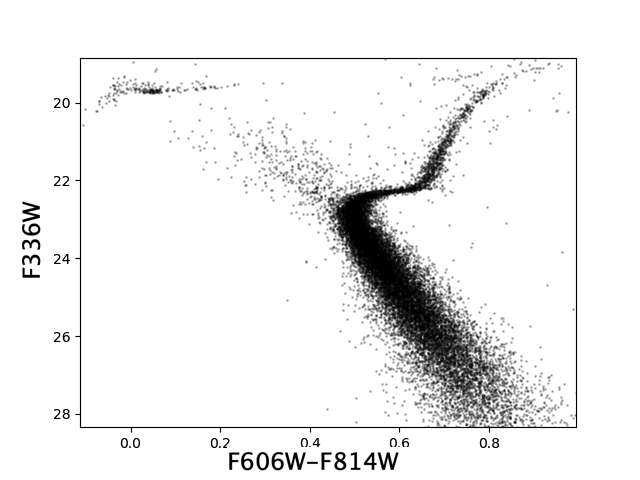}
        \\
                \small (a) &
                        \small (b)
  \end{tabular}

  \caption{Full CMDs of Hodge 11 for the F336W-F814W color (a) and the F606W-F814W color (b). These CMDs are the result of our data cleansing pipeline.}
  \label{FullCMDsHodge11}
\end{figure*}

\begin{figure*}
  \centering
  \begin{tabular}{cc}
    \includegraphics[width=\columnwidth]{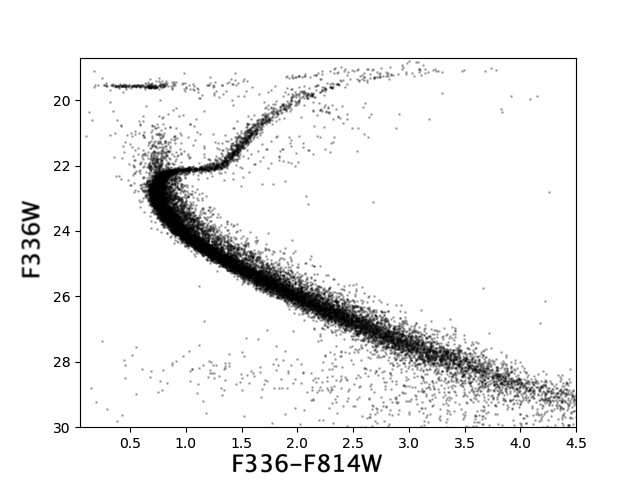} 
        \includegraphics[width=\columnwidth]{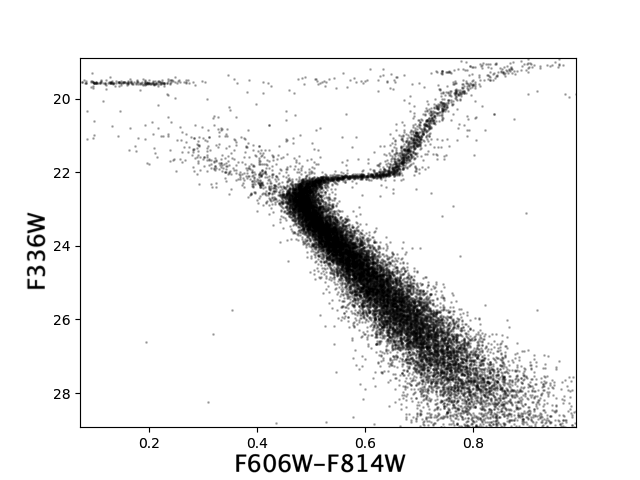}
        \\
                \small (a) &
                        \small (b)
  \end{tabular}

  \caption{Full CMDs of NGC 2210 for the F336W-F814W color (a) and the F606W-F814W color (b). These CMDs are the result of our data cleansing pipeline.}
  \label{FullCMDsNGC2210}
\end{figure*}

A complete description of the photometry and data reduction are presented in \citep{Mackeyprep}. These data were taken with the Hubble Space Telescope (GO-14164) over 54 orbits. The goal was to achieve a similar set of data for these LMC GCs as was achieved for Galactic GCs in the ACS Globular Cluster Survey \citep{Sarajedini}. Sources are reliably detected as much as 5 mag faintward of the MSTO in F336W. This can be seen in Figures \ref{FullCMDsHodge11} and \ref{FullCMDsNGC2210}. The total exposure times of each GC is on the order of 20,000 s. The photometry was reduced using Dolphot \citep{Dolphin}.

The filters used in this work are F336W, F606W, and F814W, are broadband filters that can be used to determine a GC's age and distance, and are used in techniques such as main sequence fitting using subdwarfs \citep[e.g.][]{CohenSarajedini,O'Malley}. The F336W filter is sensitive to NH and CN lines, making F336W-F606W and F336W-F814W colours potential indicators of the presence of multiple populations. However, the F336W alone is not ideal to uncover multiple populations. The F275W and F438W filters are faint in O and C rich stars while the F336W filter is bright in these stars \citep{Piotto}. The opposite is true for N rich stars. Since F606W and F814W are not strongly affected by these abundances, we only have the effects  of the multiple populations in the F336W and therefore are not able to maximize the spread in colour with our possible filter combinations. The F275W, F3336, and F438W form a `magic trio' of filters which is optimally suited to disentangle the multiple populations that are due to CNO variations \citep{MagicTrio}. However, for this work, the F275W and F438W filters were not requested because they require triple the number of orbits to achieve the same photometric quality as the selected filters.

\section{Data Analysis}

It appears that the MS of the two clusters are broadened, specifically when the F336W filter is used. This can be seen by comparing Figure \ref{FullCMDsHodge11} to Figure \ref{FullCMDsNGC2210}. To examine this further, we used a data cleansing process that follows procedures similar to those described in \citet{Milone2012}. The purpose of cleansing the data is to recover the most precise median ridge line of the MS as possible. We need a precise median ridge line for two reasons: to correct for differential reddening or un-modeled PSF variations and to straighten the colour-magnitude diagram (CMD) along the median ridge line of the MS. These straightened diagrams make it easier to visualize the presence of multiple populations.

In order to obtain the most precise median ridge line for our LMC clusters, we clean our data by keeping only the stars which have small photometric uncertainties in all three filters, F336W, F606W, and F814W. The photometric uncertainties are estimated using artificial star tests. The artificial star tests were done in a large series of runs to ensure the increased crowding did not significantly affect the photometry. It was found that Dolphot underestimates the true photometric uncertainties for most stars. These tests are described further in \citet{Mackeyprep}. We do not want to preferentially remove stars that are faint and subsequently have inherently larger errors, so the median uncertainty is calculated as a function of magnitude and stars above some multiplicative value of the median are removed. This multiplicative cut-off is individually varied for each filter in each cluster to achieve the tightest CMD, but is usually around 4 $\sigma$, though the exact cut-off used does not affect the final results. Stars near the boundary of the chip are measured fewer times and therefore are preferentially removed by this process. These methods remove around a quarter of the total number of measured stars.

We determine a median ridge line for each cluster/filter combination using a rotated histograms method first described in \citep{Marin}. The stars are divided into magnitude bins, whose width is varied from 0.1 to 0.3 mag in order to find the best fit. The overlap between the bins is 0.05 mag. The median of each bin is computed and a rotated colour-magnitude coordinate system is used so that the stellar sequence is straightened based on the initial medians.  A new set of median points in these rotated coordinates is found using the same bin parameters as above.

\begin{figure*}
  \centering
  \begin{tabular}{ccc}
    \includegraphics[width=\columnwidth]{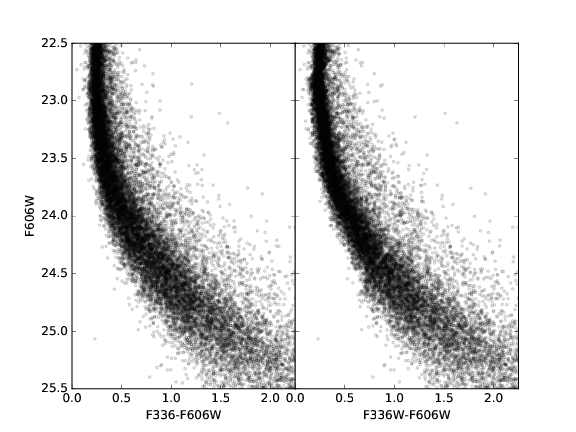} &
        \includegraphics[width=\columnwidth]{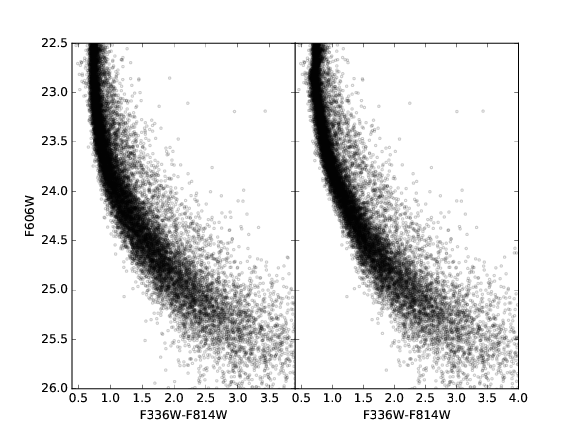}\\
        \small (a) &
                        \small (b) 
  \end{tabular}
  
  \begin{tabular}{c}
  		\includegraphics[width=\columnwidth]{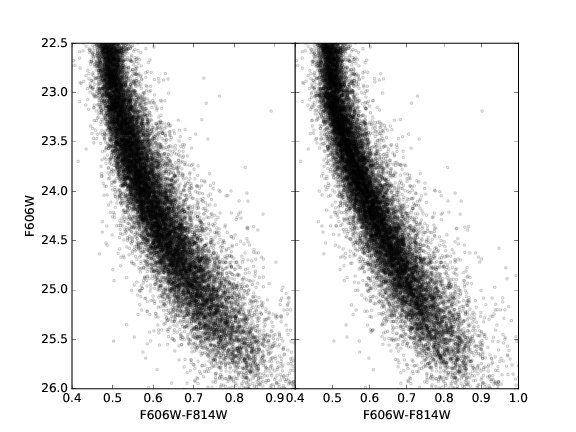} 
		\\
                 \small (c)
  \end{tabular}
  \caption{For Hodge 11, the resulting CMDs of our data cleansing pipeline. The panel on the left of a, b, and c is the before CMD and the panel on the right is the after CMD. It is clear that the CMD is noticeably tighter after removing stars with high errors and shifting stars based on expected colour as described in the text.}\label{Hodge11BeforeAfter}
\end{figure*}

As discussed in \citet{Milone2012}, geometric variations in the PSF lead to systematic errors in determining the magnitude of stars, and these errors are correlated with position. Hence, stars which are close to each other exhibit similar systematic errors. In order to correct for these systematic errors, we use the 50 closest MS stars on the chip (in other words, in pixel space) for each star and compute all of their shifts in magnitude as compared to the MS ridge line. If there is no bias on the chip nor differential reddening in the cluster, the median of all these distances should be zero. We correct the magnitude of each MS star by the median of the difference in magnitude between the 50 closest MS stars and the ridge line. This is repeated for all three colours.  The full, cleansed, CMDs for Hodge 11 and NGC 2210 are shown in Figures \ref{FullCMDsHodge11} and \ref{FullCMDsNGC2210}. To more easily see the effect of the data cleansing process, the initial and corrected MS of Hodge 11 is shown in Figure \ref{Hodge11BeforeAfter}. The left panel is the initial MS and the right panel is the cleaned MS. 

\begin{figure*}
  \centering
  \begin{tabular}{cc}
    \includegraphics[width=\columnwidth]{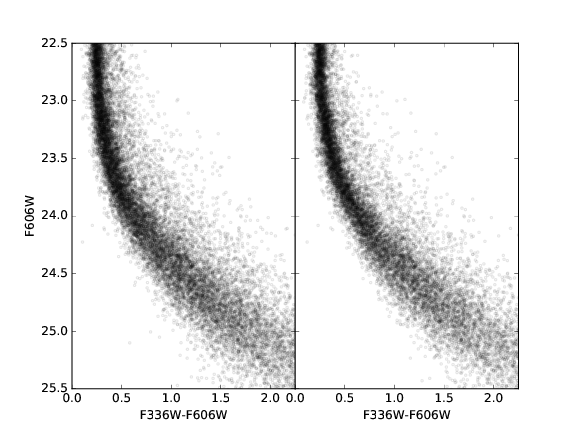} &
        \includegraphics[width=\columnwidth]{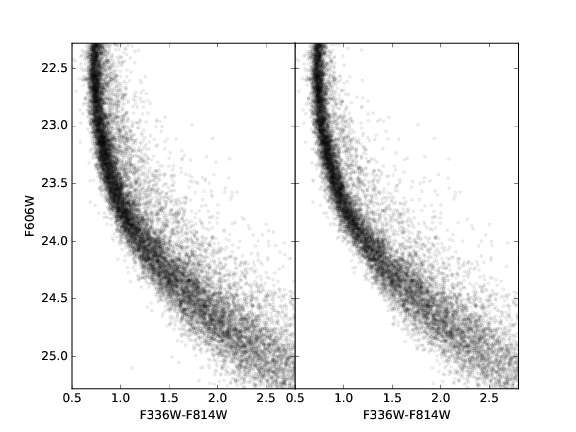} \\
        \small (a) &
                        \small (b) \\

  \end{tabular}
  
    \begin{tabular}{cc}
     		\includegraphics[width=\columnwidth]{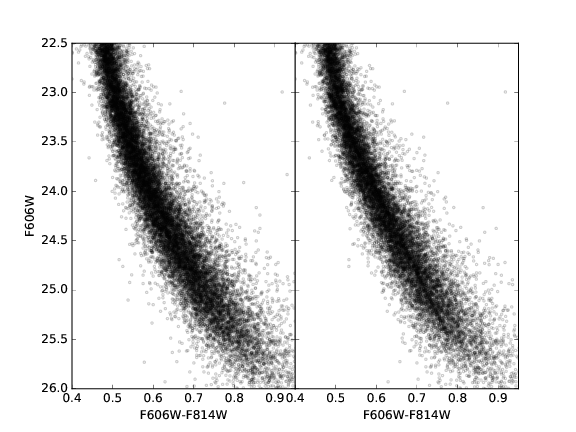}
		\\
                \small (c)
  \end{tabular}
  
  \caption{For NGC 2210, the resulting CMDs of our data cleansing pipeline. The panel on the left of a, b, and c is the before CMD and the panel on the right is the after CMD. It is clear that the CMD is noticeably tighter after removing stars with high errors and shifting stars based on expected colour as described in the text.}\label{NGC2210BeforeAfter}
\end{figure*}

\begin{figure*}
  \centering
  \begin{tabular}{cc}
    \includegraphics[width=\columnwidth]{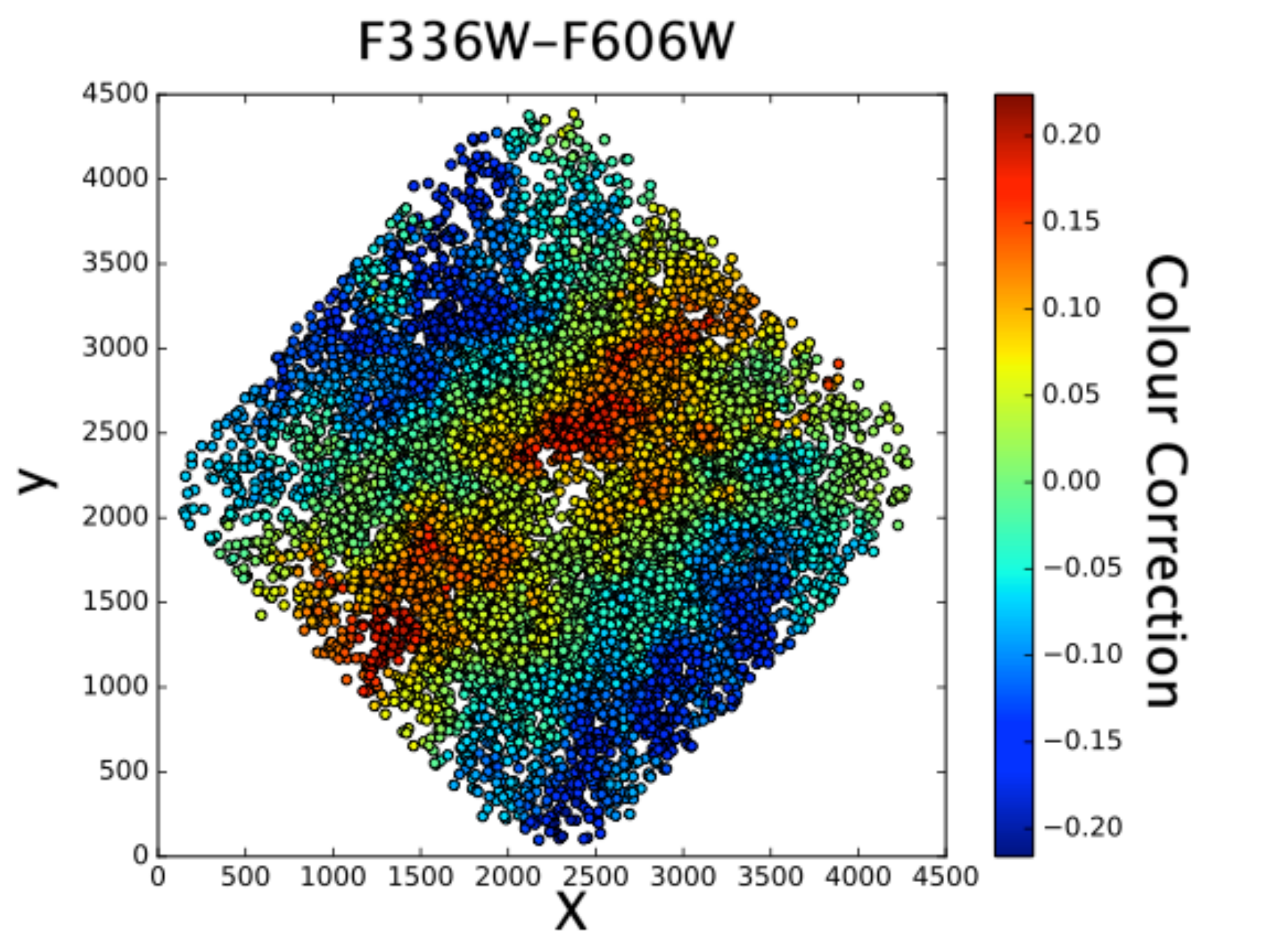} &
        \includegraphics[width=\columnwidth]{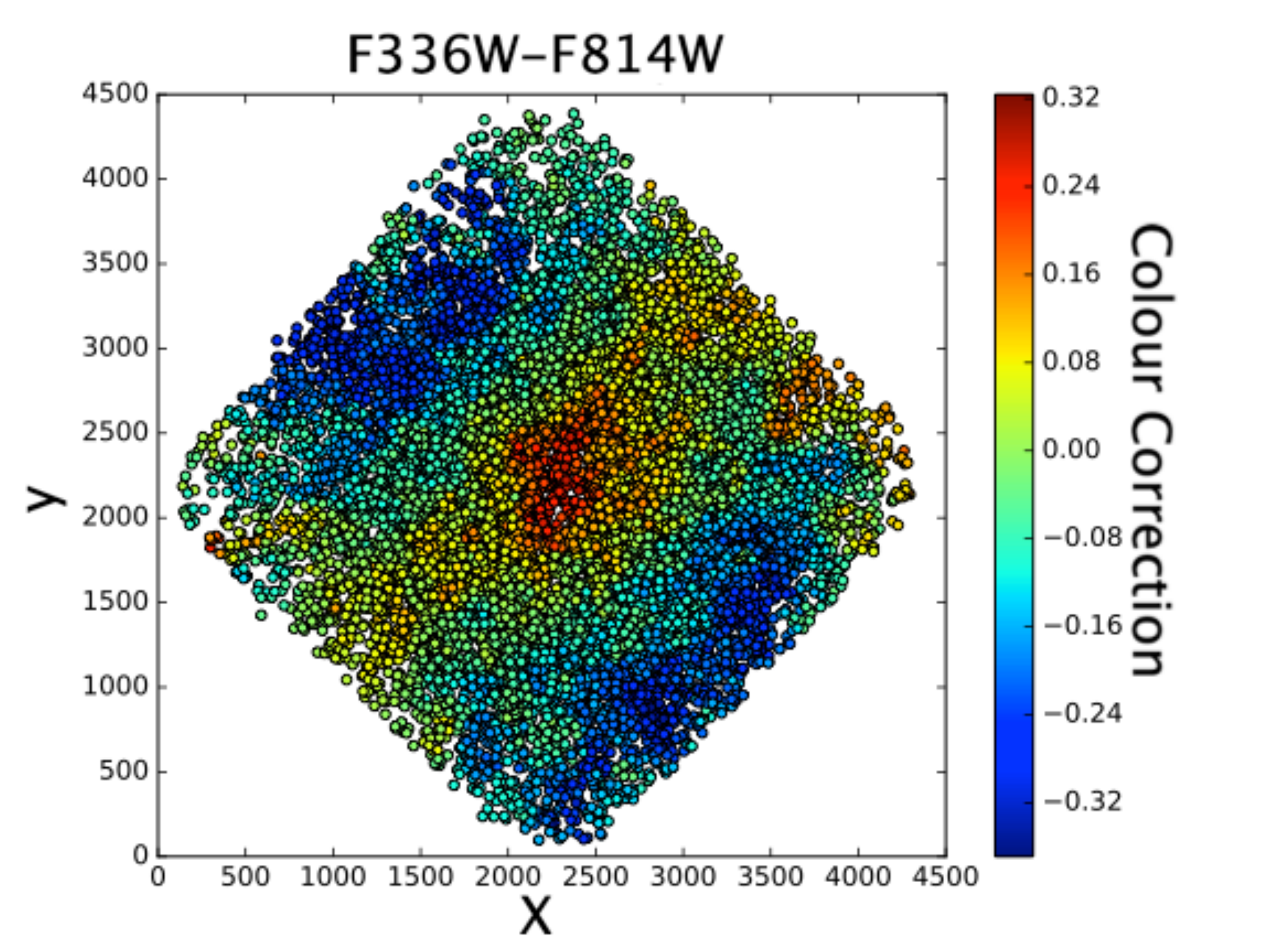} 
		\\
                \small (a) &
                        \small (b)
  \end{tabular}
    \begin{tabular}{c}
 		\includegraphics[width=\columnwidth]{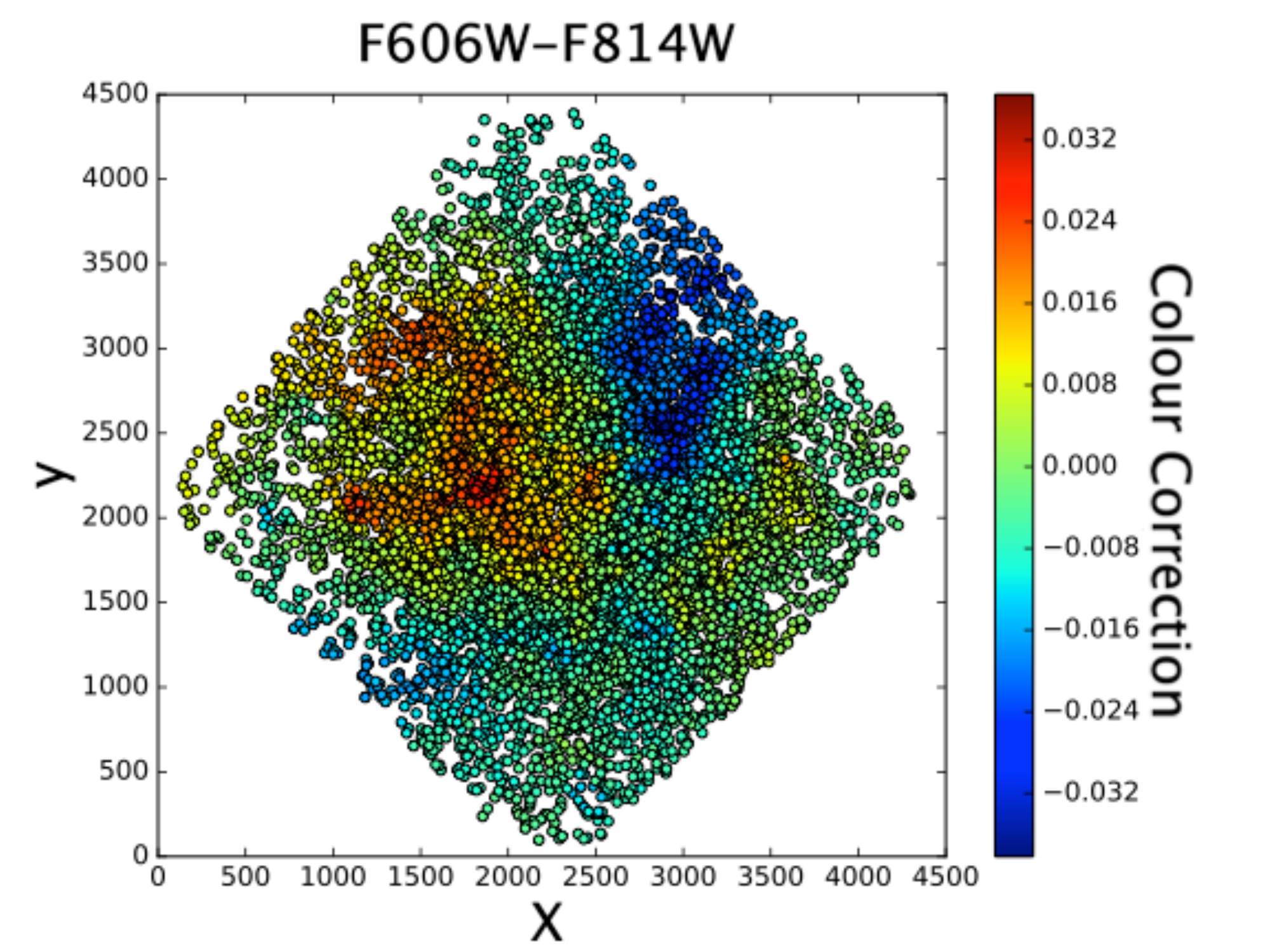}
 		\\
 		\small (c)
 \end{tabular}
  \caption{For Hodge 11, the colour shift applied to each star. It is calculated by taking the 50 nearest stars' distances to the median ridge line which is then applied to the star.}
\label{Hodge11ColorShift}
\end{figure*}

\begin{figure*}
  \centering
  \begin{tabular}{cc}
    \includegraphics[width=\columnwidth]{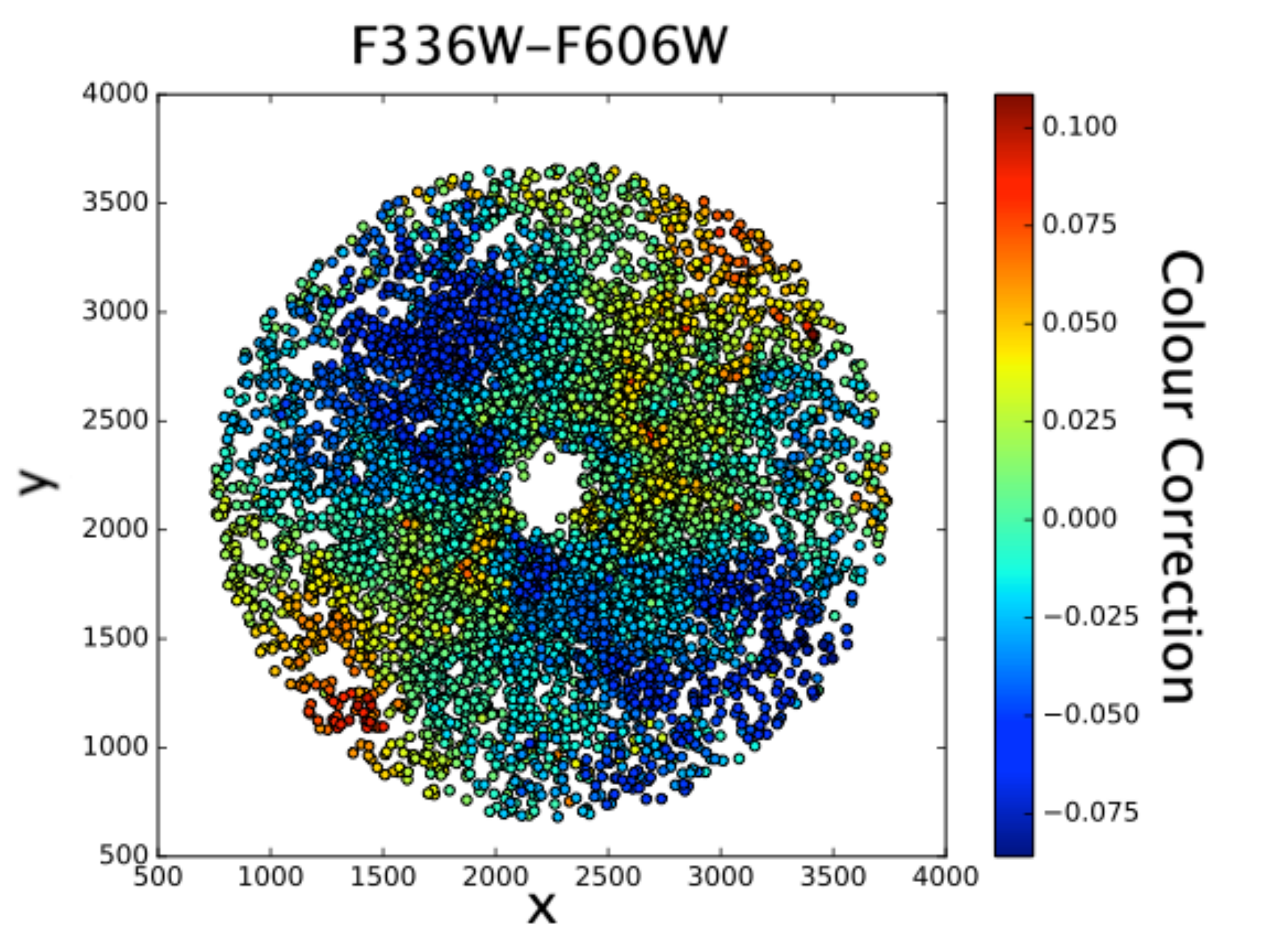} &
        \includegraphics[width=\columnwidth]{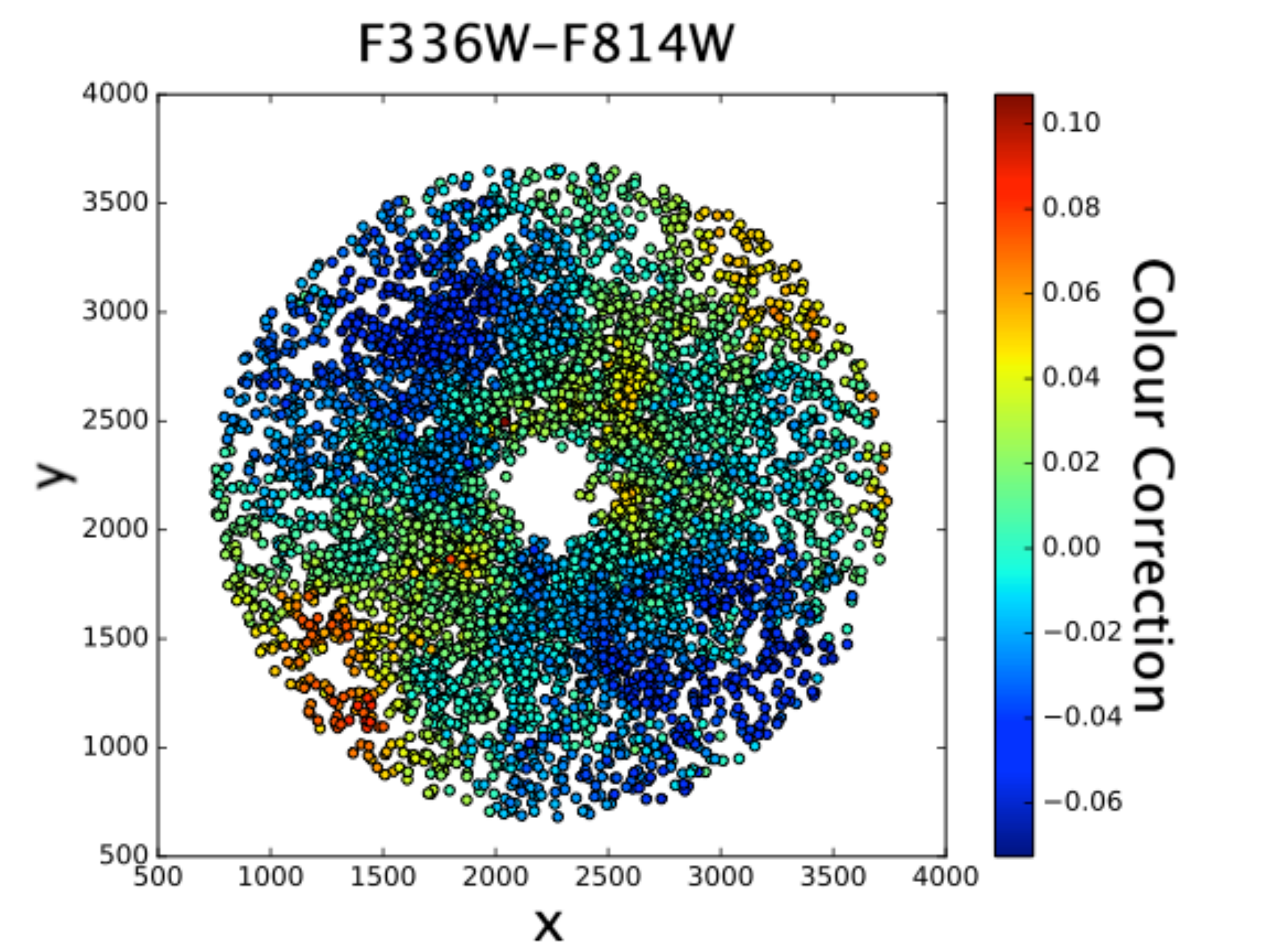}
		\\
                \small (a) &
                        \small (b)
  \end{tabular}
  
    \begin{tabular}{c}
     \includegraphics[width=\columnwidth]{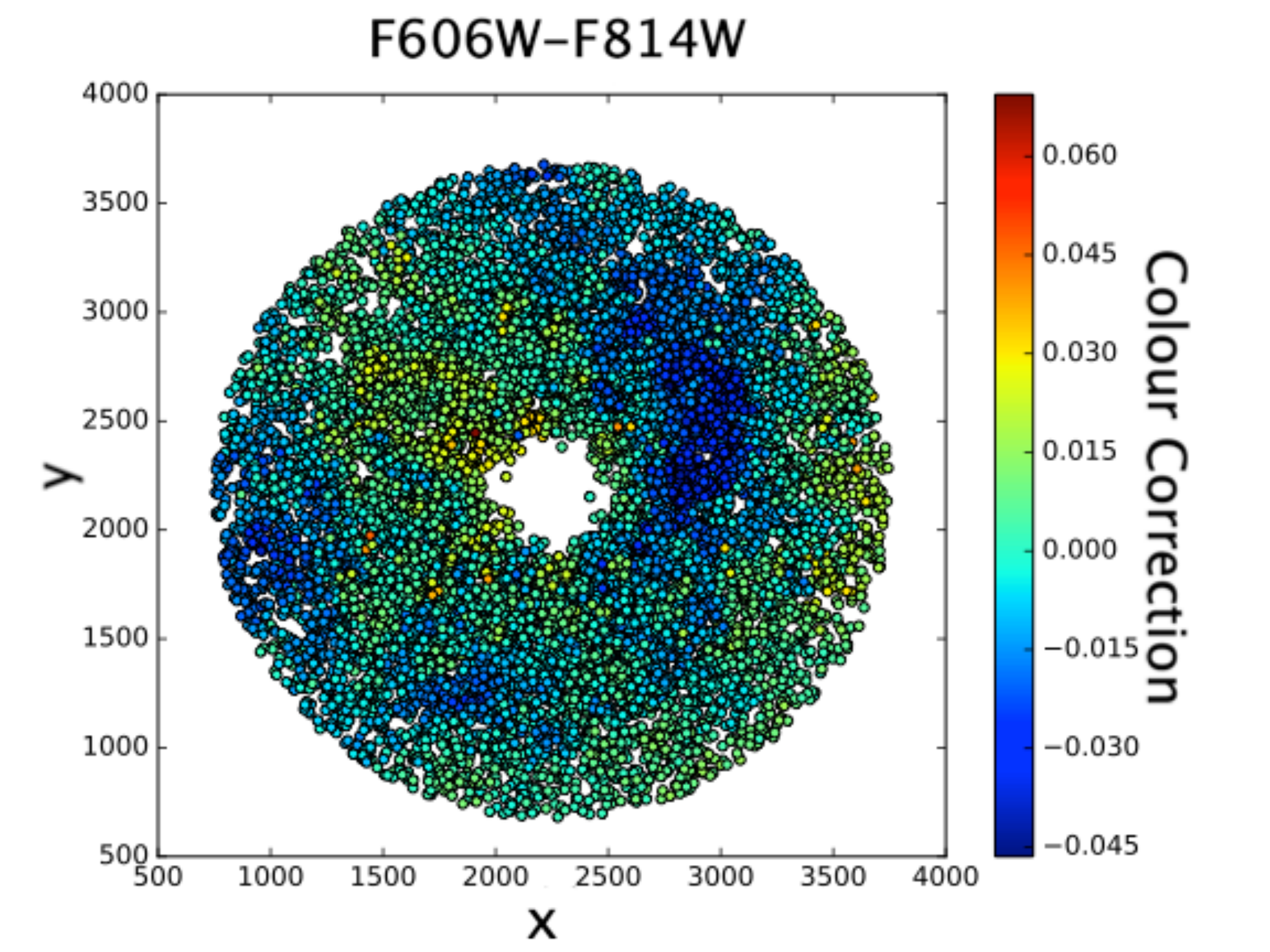} 
     \\
     \small (c)
  \end{tabular}
  \caption{For NGC 2210, the colour shift applied to each star. This is with a 2000 radial pixel cut. It is calculated by taking the 50 nearest stars' distances to the median ridge line which is then applied to the star. Using a similar cut on Hodge 11 did not change the results and therefore are not performed.}
\label{NGC2210ColorShift}
\end{figure*}

The data are noticeably tighter after our corrections. The result for NGC 2210 is Figure \ref{NGC2210BeforeAfter}. Figures \ref{Hodge11ColorShift} and \ref{NGC2210ColorShift} show the magnitude of these colour shifts as a function of chip position. While the largest changes are on the order of $\pm0.1$ mag, the median and average of these changes are only around $\pm0.025$ mag.

\begin{figure*}
  \centering
  \begin{tabular}{cc}
    \includegraphics[width=\columnwidth]{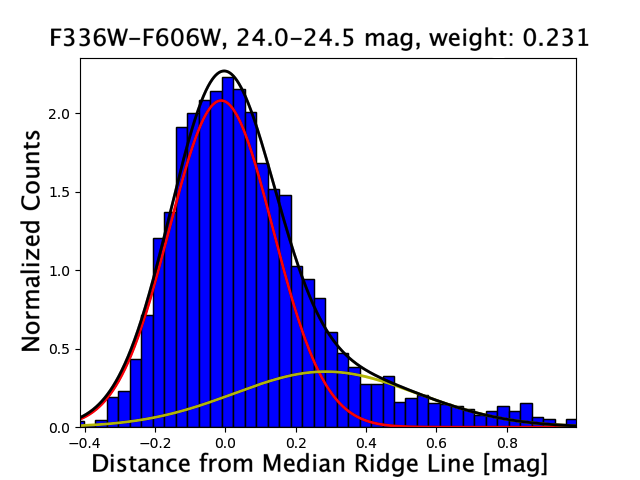} &
        \includegraphics[width=\columnwidth]{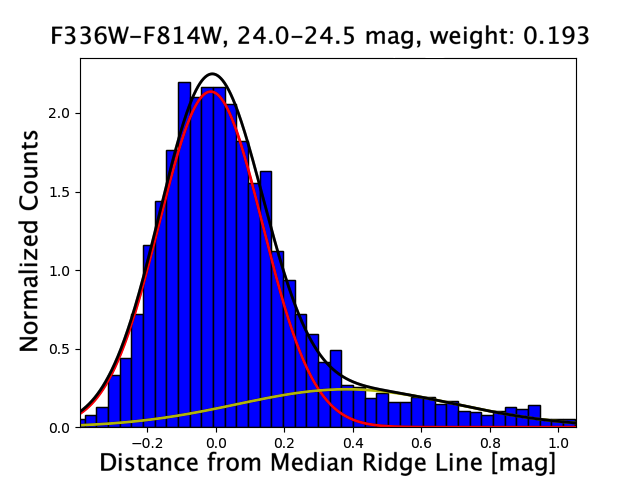} 
		\\
                \small (a) &
                        \small (b)
  \end{tabular}
  
  \begin{tabular}{c}
 		\includegraphics[width=\columnwidth]{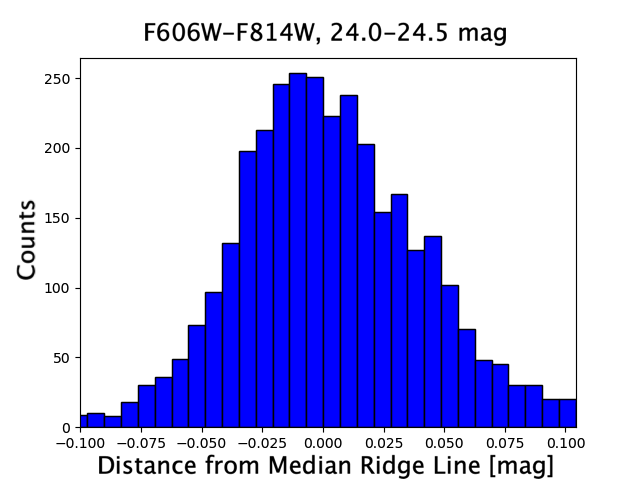}
 		\small (c)
  \end{tabular}
  \caption{For Hodge 11, a selection of histograms. There is a clear red tail in the F336W-F606W and F336W-F814W colours. It is not present in the F606W-F814W colour. The weight parameter is described in Equation \ref{DoubleGauss}.}
\label{Hodge11Histograms}
\end{figure*}

\begin{figure*}
  \centering
  \begin{tabular}{cc}
    \includegraphics[width=\columnwidth]{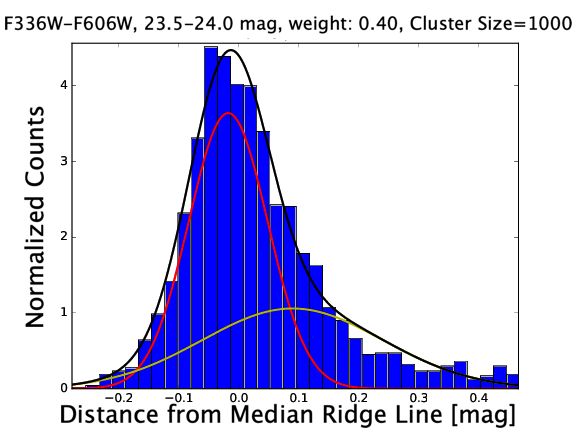} &
        \includegraphics[width=\columnwidth]{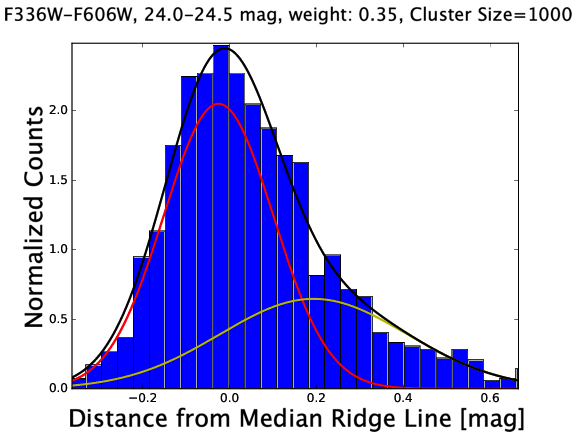}\\
         \small (a) &
                        \small (b)\\
 		\includegraphics[width=\columnwidth]{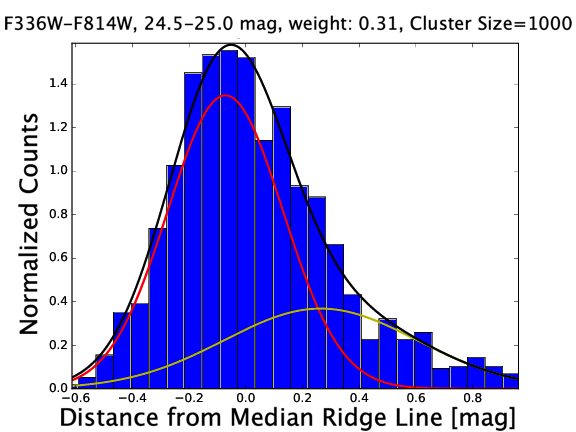} &
 		 \includegraphics[width=\columnwidth]{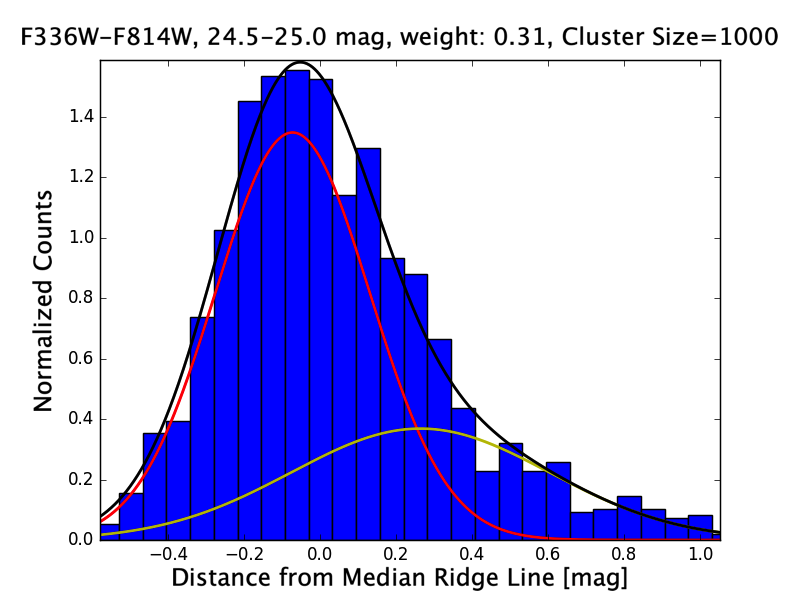}\\
 		                \small (c) & \small (d)\\
 		\end{tabular}
      \begin{tabular}{c}
        \includegraphics[width=\columnwidth]{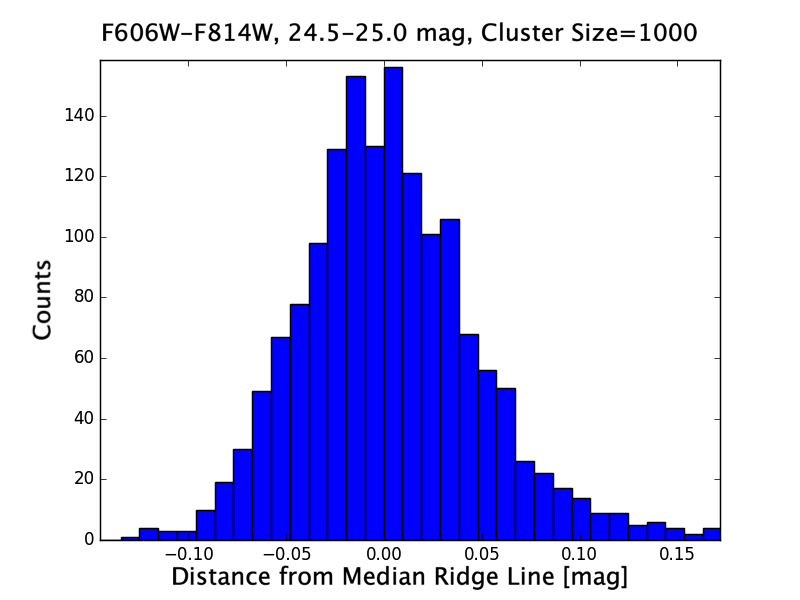}
		\\
		 \small (e)
  \end{tabular}

  \caption{For NGC 2210, a selection of histograms with a cluster size of 1000 pixels. There is a clear red tail in the F336W-F606W and F336W-F814W colours. It is not present in the F606W-F814W colour. The weight parameter is described in Equation \ref{DoubleGauss} and estimates the size of the second population.}
\label{NGC2210Histograms1000}
\end{figure*}

We wish to test whether a multiple populations scenario can provide a good description of the data, so we fit two populations to the cleaned data using a linear least squares method. This is shown for each cluster in Figures \ref{Hodge11Histograms} and \ref{NGC2210Histograms1000} for several magnitude bins. The weight parameter, $w$, is given by Equation \ref{DoubleGauss}.

\begin{figure*}
  \centering
  \begin{tabular}{cc}
    \includegraphics[width=\columnwidth]{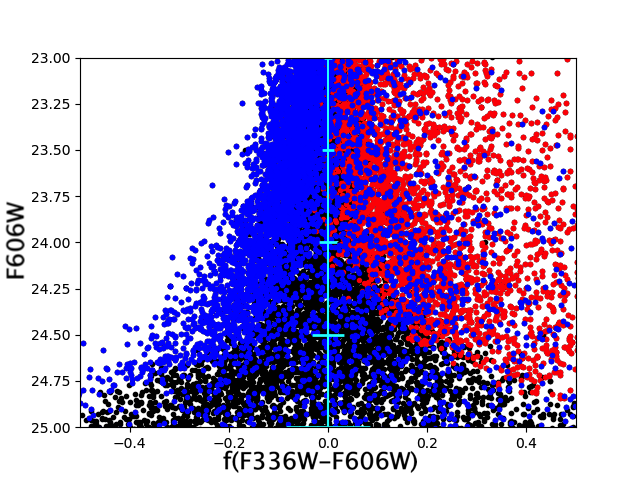} &
        \includegraphics[width=\columnwidth]{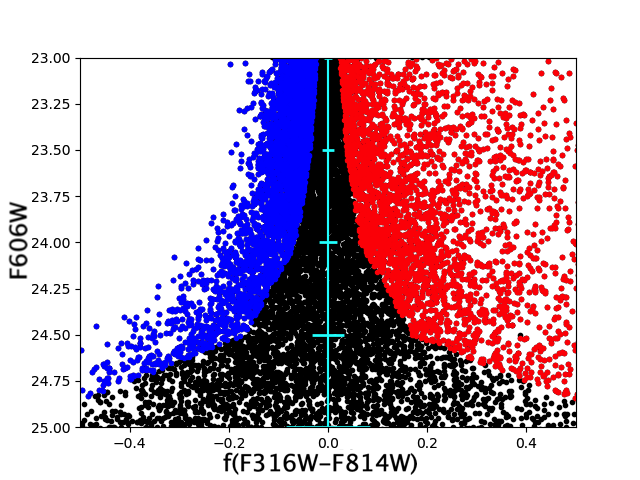}

		\\
                \small (a) &
                        \small (b)
  \end{tabular}
    \begin{tabular}{c}
 		\includegraphics[width=\columnwidth]{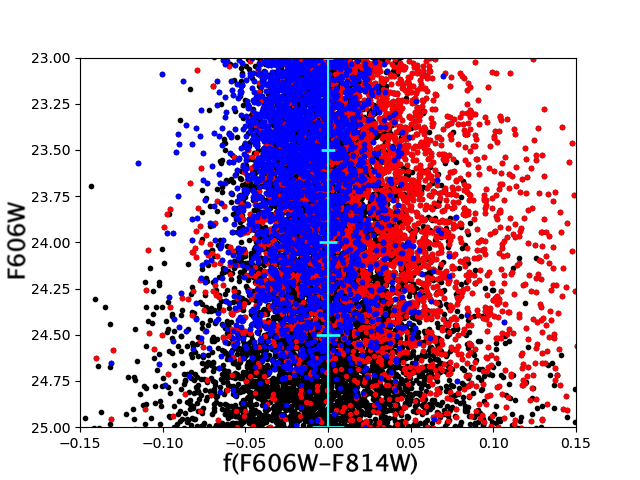} \\
 		\small (c)
 		  \end{tabular}
  \caption{For Hodge 11, red and blue stars. The red and blue stars were defined from the f(336-814) versus F606W straightened CMD. We plot these same red and blue in the other panels. This is a check of binary contamination. Binaries would appear redder in every colour. Clearly there is some binary contamination as expected, but this is not the only cause of the excess of red stars.}
\label{Hodge11RedBlue}
\end{figure*}

\begin{figure*}
  \centering
  \begin{tabular}{cc}
    \includegraphics[width=\columnwidth]{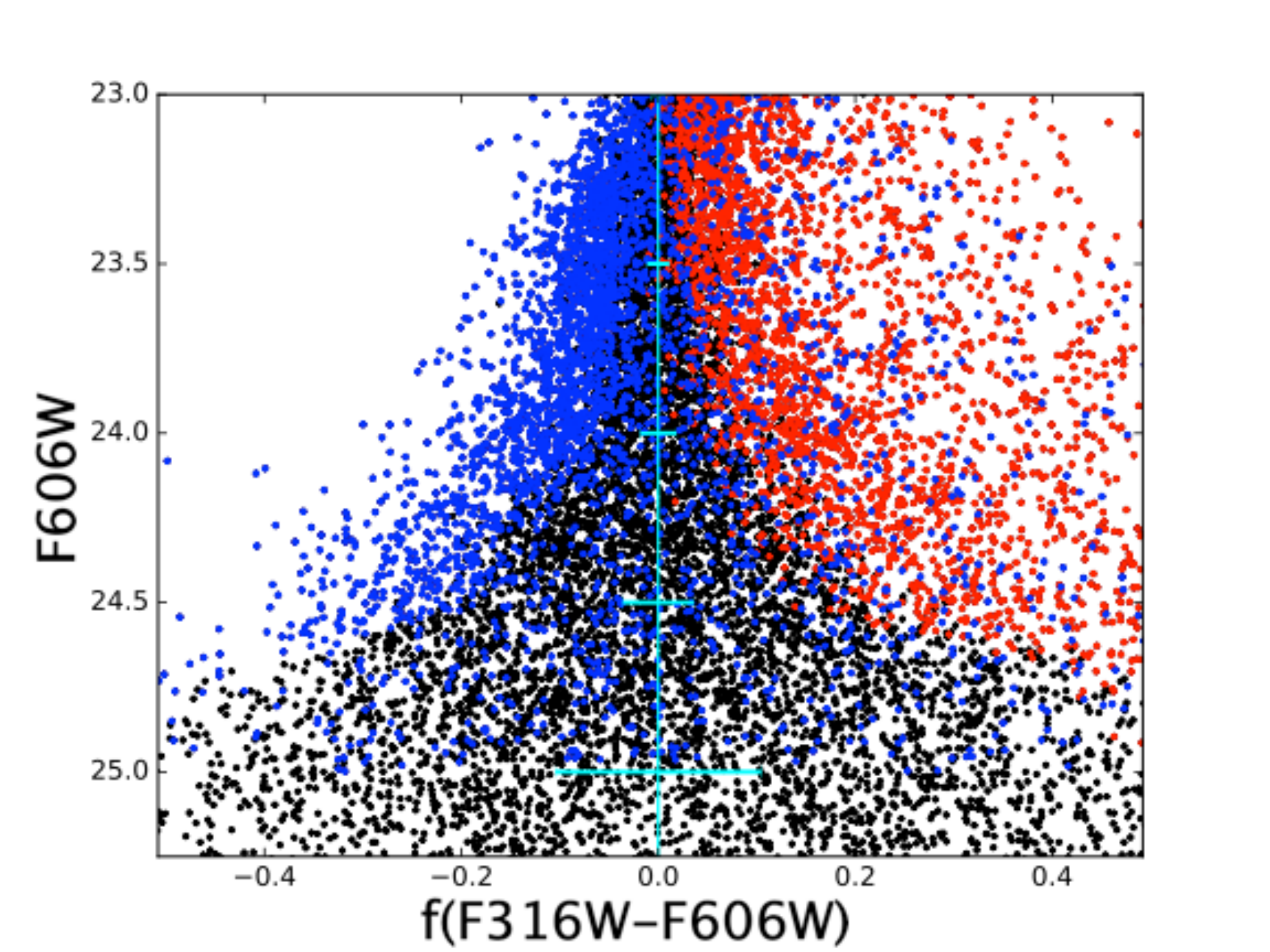} &
        \includegraphics[width=\columnwidth]{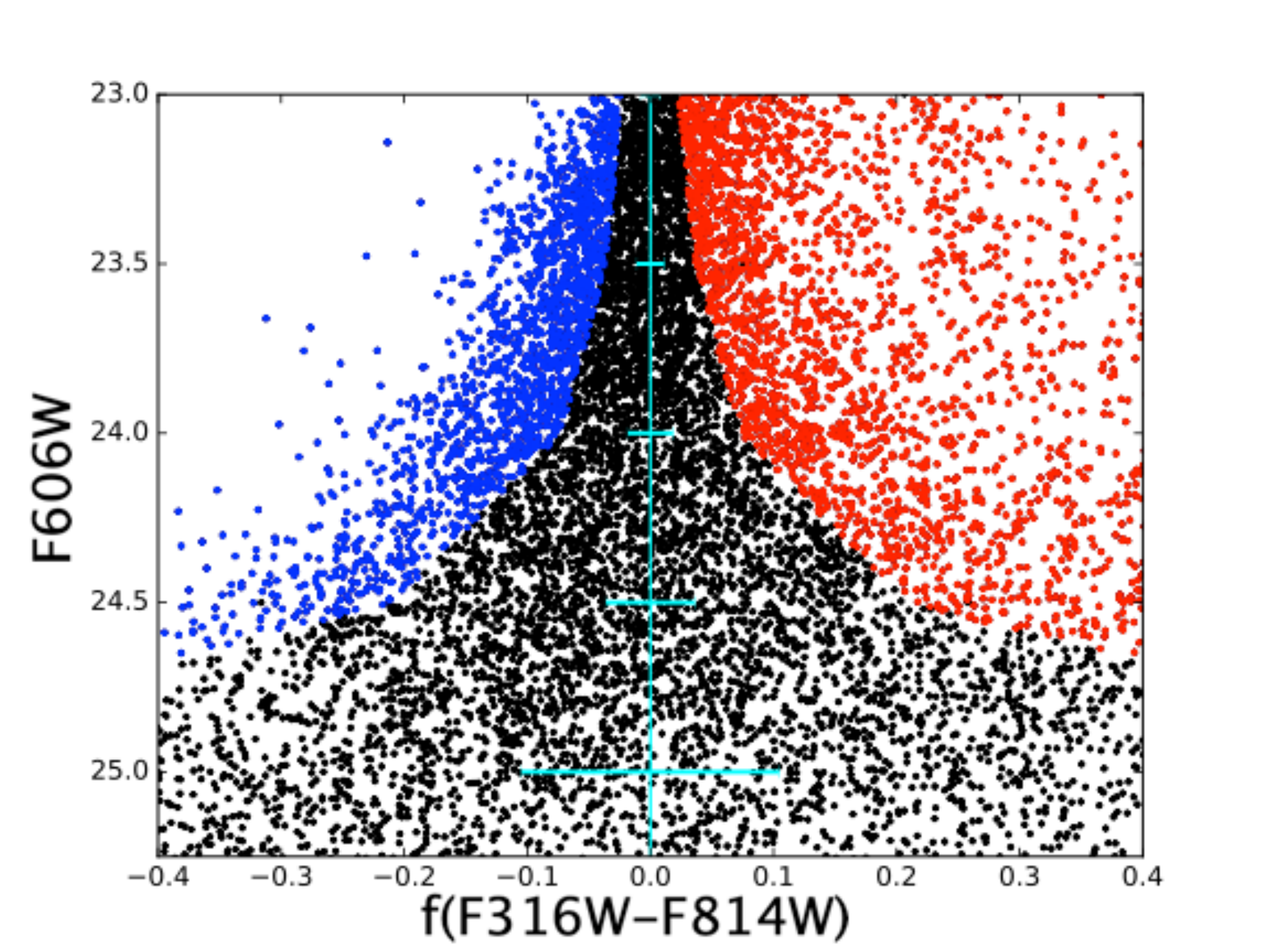}
 	
		\\
                \small (a) &
                        \small (b)
  \end{tabular}
    \begin{tabular}{c}
  	\includegraphics[width=\columnwidth]{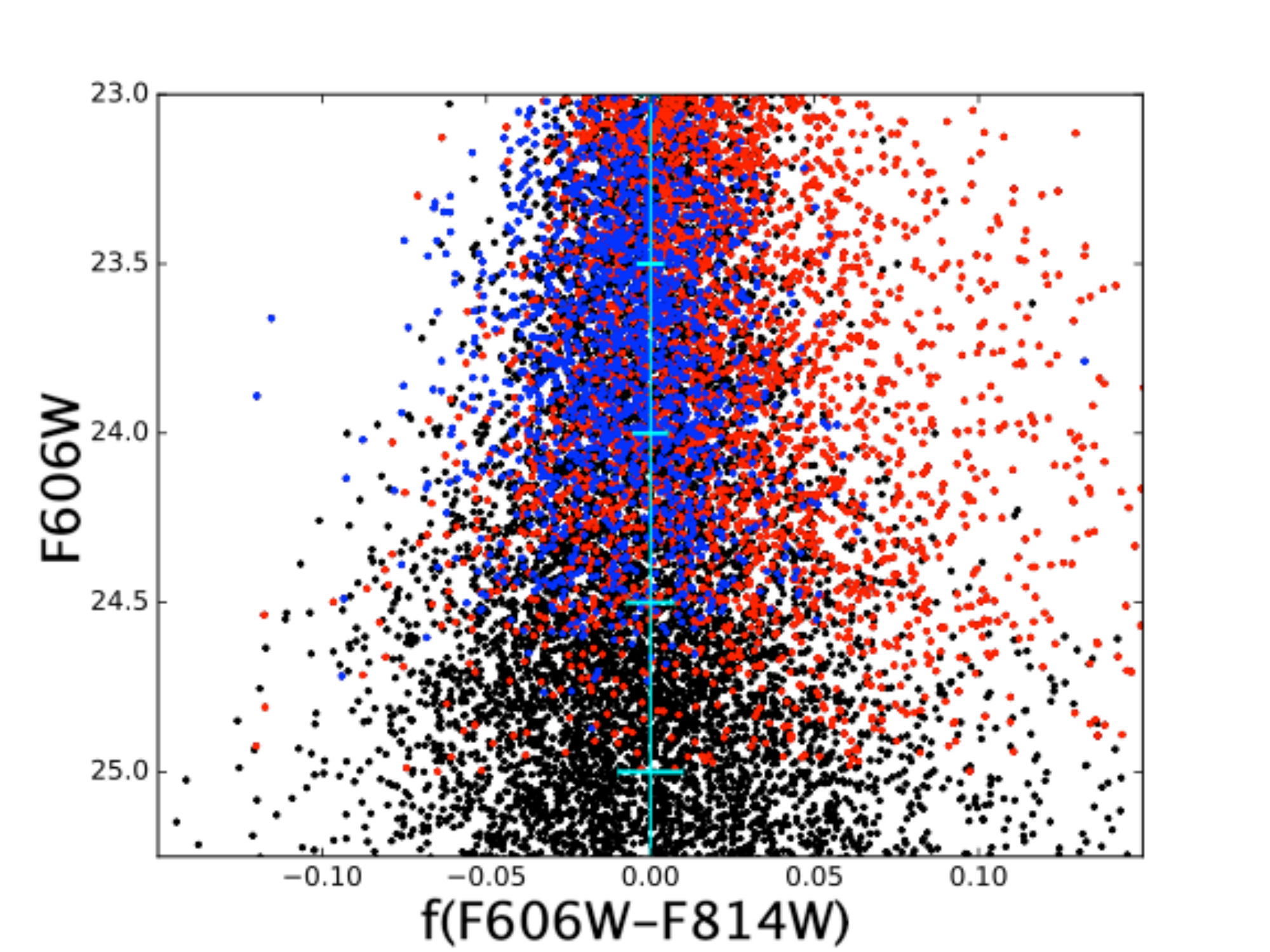} \\
  	\small (c)
  \end{tabular}
  \caption{For NGC 2210, red and blue stars. The red and blue stars were defined from the f(336-814) versus F606W straightened CMD. We plot these same red and blue in the other panels. This is a check of binary contamination. Binaries would appear redder in every colour. Clearly there is some binary contamination as expected, but this is not the only cause of the excess of red stars.}
\label{NGC2210RedBlue}
\end{figure*}
 
\begin{figure*}
  \centering
  \begin{tabular}{cc}
    \includegraphics[width=\columnwidth]{Hodge11Histogram3.png} &
        \includegraphics[width=\columnwidth]{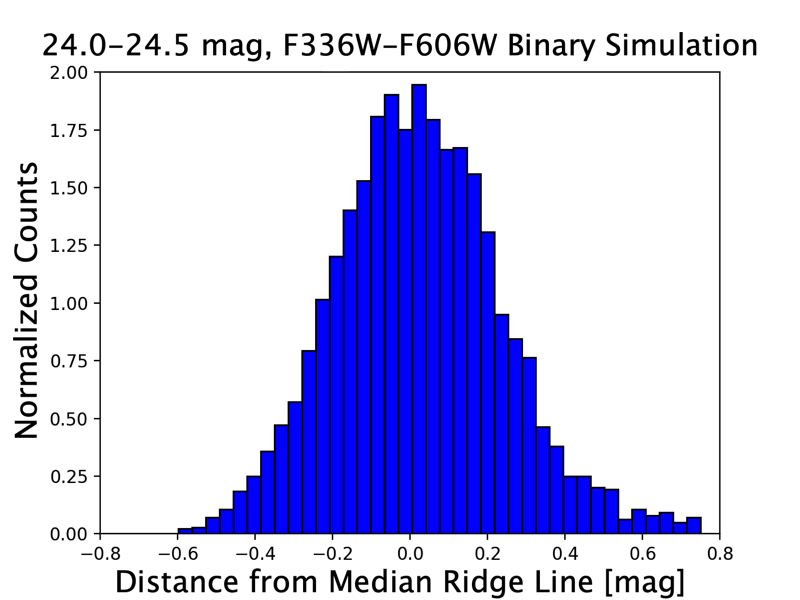} 
		\\
                \small (a) &
                        \small (b)
  \end{tabular}

  \caption{Binary simulation results for Hodge 11. The true histogram for Hodge 11 is shown in Panel (a). Using the primordial sequence, we create a simulated histogram in Panel (b). We also add in binary star contamination. Comparing Panels (a) and (b), it is clear that adding in binaries does not create a distribution that is similar to the data.}
     \label{Binary}
\end{figure*}

\begin{figure*}
  \centering
  \begin{tabular}{cc}
    \includegraphics[width=\columnwidth]{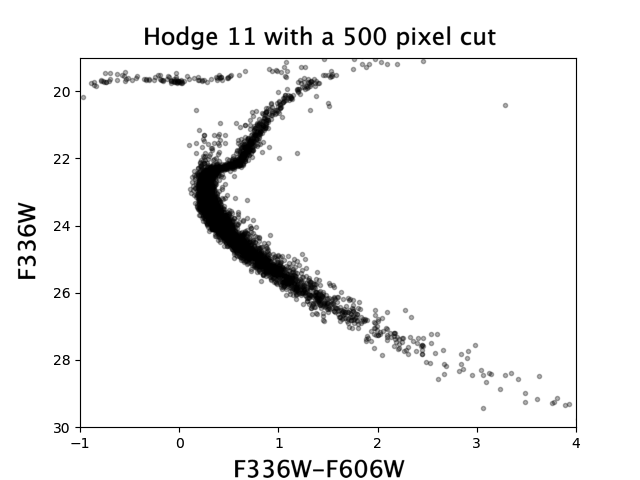} &
        \includegraphics[width=\columnwidth]{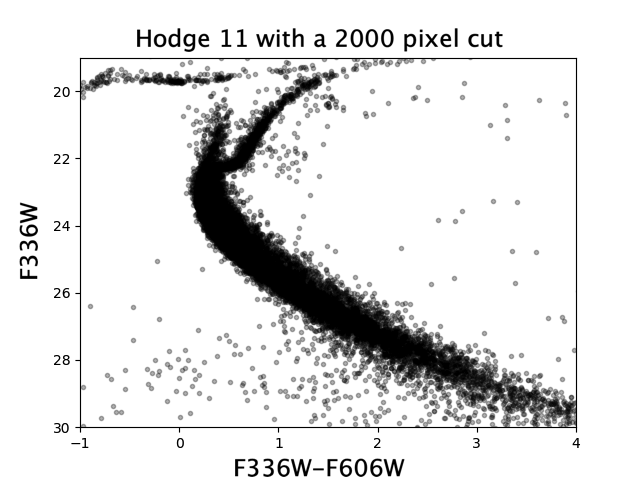} 
		\\
                \small (a) &
                        \small (b)
  \end{tabular}

  \caption{Panel (a) shows the CMD for Hodge 11 for a 500 pixel radial cut while Panel (b) shows the same CMD but with a 2000 pixel cut.}
     \label{Hodge11RadialCut}
\end{figure*}

\begin{equation}
\label{DoubleGauss}
f=(1-w)\exp\left(-\frac{(x_{1}-\mu_{1})^{2}}{2\sigma_{1}^{2}}\right)+w\exp\left(-\frac{(x_{2}-\mu_{2})^{2}}{2\sigma_{2}^{2}}\right)
\end{equation}

where $x$ is the center, $\mu$ is the mean and $\sigma$ is the standard deviation of each Gaussian population.

The weight factor's precise value is somewhat uncertain. When the two populations' means ($\mu_{1}$ and $\mu_{2}$)  are too close, the weight factor has some degeneracy in the fitting. Therefore, the weight factor is not well constrained by our data, but is indicative of a relative size of the second population. 

The two-population Kolmogorov-Smirnov (KS) test is a statistical measure of the likelihood that two sets of data are drawn from the same population. We create a Monte Carlo simulation of a set of stars with the same luminosity function as the cluster over a grid of parameters ($\mu$, $\sigma$, $x_{2}$). We use the two-population KS test to compare the model with the data. The model with the highest confidence of being drawn from the same population is nearly the same as derived using the linear-least squares method. The confidence for these parameters is nearly always greater than 99\%.

Besides multiple populations, there are two other possible reasons that this red population exists: photometric errors and binary systems. To test if the spread is due to binaries, we select stars that are much redder or bluer than the median in the f(336-814) straightened CMD and colour code these stars as red or blue in all three diagrams, shown Figures \ref{Hodge11RedBlue} and \ref{NGC2210RedBlue}. We expect that binary star systems would appear to be redder in all colours since the luminosity is increased but the colour is only weakly, if at all, changed. We adopt a definition of a binary star to be brighter than the median ridge line by at least 0.35 mag in all three colour-magnitude combinations. This technique for disentangling a redder secondary population from the binaries was developed for NGC 1851 \citep{Cummings2014}. 10\% of the main sequence stars in Hodge 11 are in binary systems and 12\% are for NGC 2210 using this parameter. This number does not fully account for the extended red sequence exhibited by both clusters. 

Because it is hard to disentangle binaries with small mass ratios from a single star, we create a Monte Carlo simulation in order to see whether this extended main sequence is solely due to binary star systems. Using \citet{Binaries} as a reference, the simulation creates a set of stars with the same luminosity function as the two clusters. We then add in binary star systems that follow a flat distribution in colour up to 0.7 mag brighter than the median ridge line. The result for one magnitude bin is shown in Figure \ref{Binary}. None of the histograms created in this simulation match the features found for these two clusters. The tails of the distributions are smaller in amplitude and less extended than our data and are better fitted using a one population model than a two population model. A two-population KS test is run, showing that the binary simulation model and our data are not drawn from the same distributions with a greater than 98\% certainty.

An open question about GC populations is the radial distribution of the first and subsequent populations. Formation scenarios that involve massive stars, either rotating massive stars or AGB stars, are believed to produce a centrally located second population compared to the original \citep{Larsen}. One reason is due to mass-segregation in GC. Some work, including \citet{Milone2012} and \citet{Larsen}, supports this assumption and find that the second population is centrally located. However, in an analysis of NGC 6362, \citet{Dalessandro} find that the primordial population can be located more centrally than the second population. However, due to the mixing time scales of GCs being relatively short, it is possible that after 1-2 Gyr the two populations are fully mixed.  

With this in mind, we examine each cluster with varying radial cuts. We examine their CMDs and histograms with respect to the radial cuts. The F336W-F606W radial cut is shown for Hodge 11 in Figure \ref{Hodge11RadialCut}. The first panel is with a cut of within 500 pixels of the centre of the cluster while the second panel is with a 2000 pixel cut. The stars that were removed in the data cleansing process are also not included. It is clear that there are many more field stars when stars further from the centre of the cluster are included. Many apparent blue straggler stars are mainly seen on the outskirts of the cluster. These stars could be true blue stragglers, LMC field stars, or most unlikely, MW foreground stars.

\begin{figure*}
  \centering
  \begin{tabular}{cc}
    \includegraphics[width=\columnwidth]{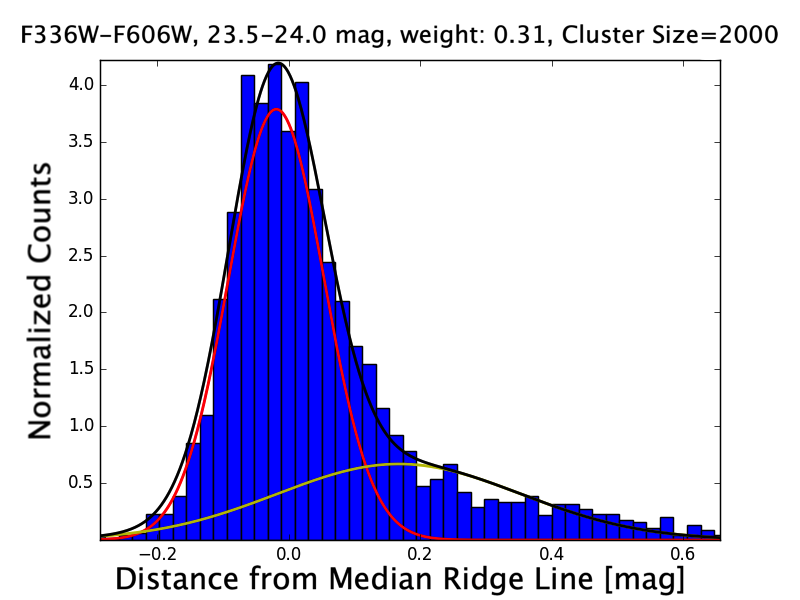} &
        \includegraphics[width=\columnwidth]{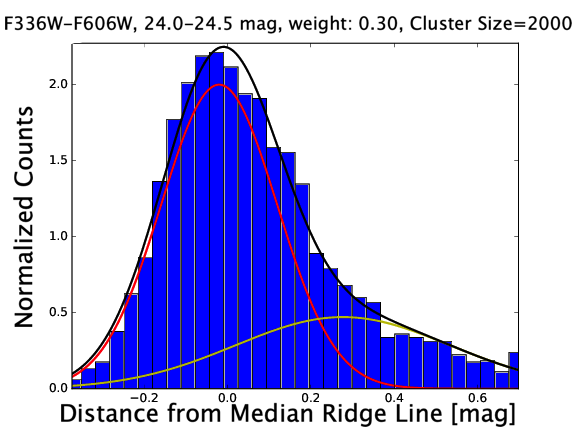}\\
         \small (a) &
                        \small (b) \\
 		\includegraphics[width=\columnwidth]{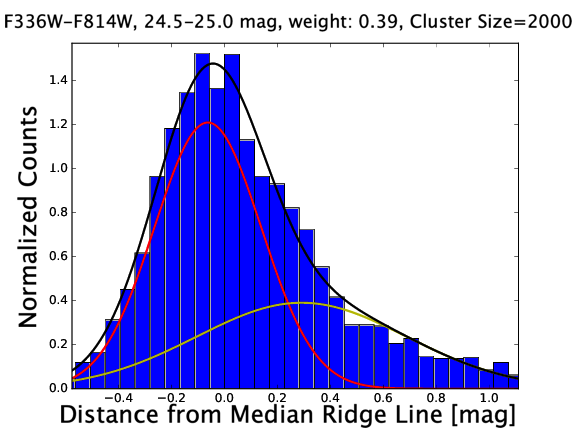} &
    \includegraphics[width=\columnwidth]{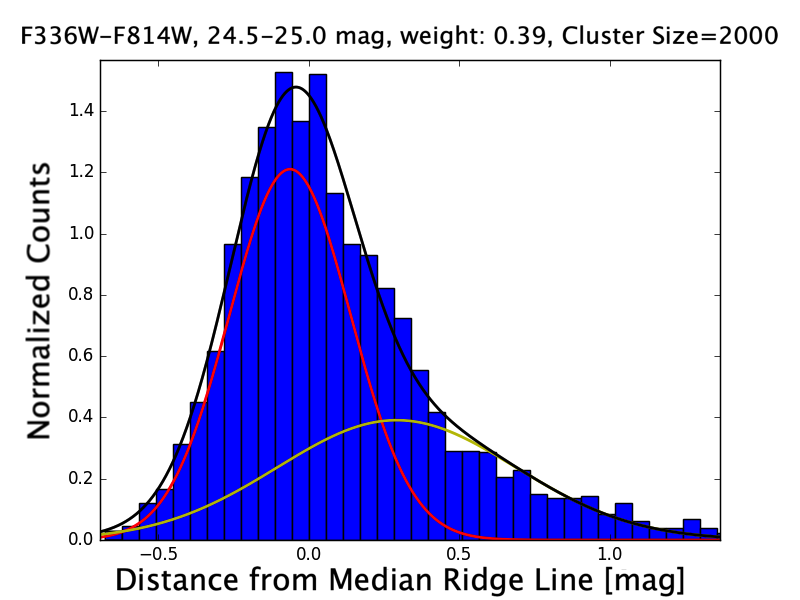}\\
     		                \small (c) & \small (d)
     
  \end{tabular}		                
       \begin{tabular}{c}		                
        \includegraphics[width=\columnwidth]{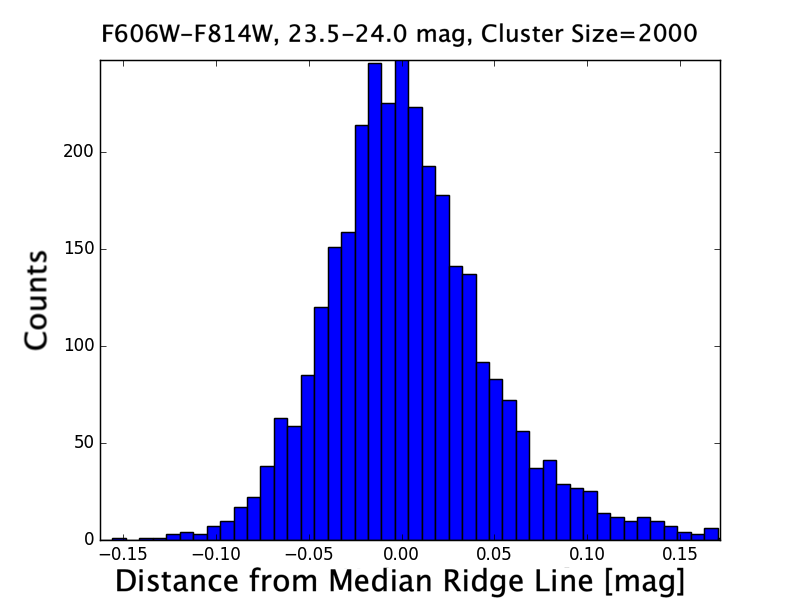}
		\\
		\small (e)

  \end{tabular}

  \caption{For NGC 2210, a selection of histograms with a cluster size of 2000 pixels. There is a clear red tail in the F336W-F606W and F336W-F814W colours. It is not present in the F606W-F814W colour. The weight parameter is described in Equation \ref{DoubleGauss}.}
\label{NGC2210Histograms2000}
\end{figure*}

A radial cut based on cluster position is only significant for histograms of NGC 2210; including this cut for the Hodge 11 histograms had no measurable effect. The purpose of this cut is to remove field stars and other non-cluster members. Stars toward the higher stellar density cluster centre were preferentially removed by the data cleansing pipeline causing the ``hole" in Figure \ref{NGC2210ColorShift}. Overall, increasing the cluster size had an effect of up to 9\% difference in the size of the secondary population for a given colour and magnitude bin as shown in Figure \ref{NGC2210Histograms2000}. However, the overall size of the secondary population using all of the various histograms stays in the same range. The errors of the secondary population of NGC 2210 includes the uncertainty in the size of the secondary population due to positional effects. For Hodge 11, there was no difference in population size based on radial cut so it does not affect the overall uncertainty. For F336W-F606W, this increase caused the weight of the second population to fall. However, for F336W-F814W, the second population's weight actually increased.

\begin{figure*}
  \centering
  \begin{tabular}{cc}
    \includegraphics[width=\columnwidth]{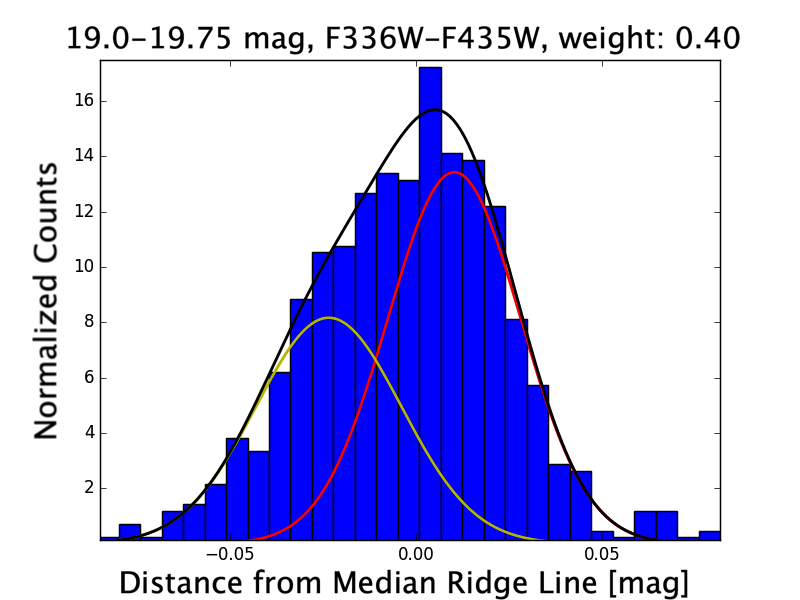} &
        \includegraphics[width=\columnwidth]{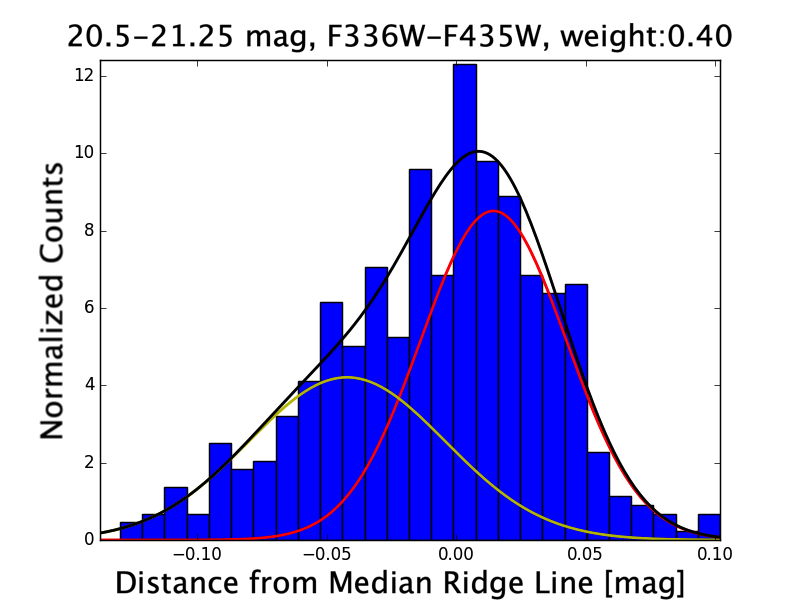} 
		\\
                \small (a) &
                        \small (b)
  \end{tabular}

  \caption{Results of the degradation of the photometry of NGC 6397. These histograms are an analogue to Figure 8 from \citet{NGC6397}. The yellow Gaussian is the primoridal population while the red Gaussian is the second generation.}
     \label{NGC6397Hist}
\end{figure*}

To check the robustness of our data cleansing process, we repeat this analysis with other clusters examined previously and compared the results. NGC 6397 has a split main sequence \citep{NGC6397}. Running this cluster's photometry through our procedure produces results that match what is demonstrated in \citet{NGC6397}. We then apply a degradation of the data so that the errors are similar to our data. Errors from both NGC 2210 and Hodge 11 are used to degrade the photometry, but since their photometric errors are similar, we find no difference between the two in the final result. Figure \ref{NGC6397Hist} is an analogue of the sidebar of Figure 8 from \citet{NGC6397}. The peak of the second population is less defined than in the original photometry. The second component in the original work is around 29\% $\pm$ 3\% of the total population while our reduction puts it at 40\%.

\section{Main Sequence Results}

Both NGC 2210 and Hodge 11 main sequences show evidence for multiple populations of stars. The other population is redder for both clusters. The effects of multiple chemical abundances on isochrones are difficult to predict without the use of stellar models and are only as accurate as these stellar models. An obvious effect is that by enhancing a star with helium and therefore increasing the star's mean molecular mass, the hydrogen burning rates are increased  \citep{Salaris}. Adding to this effect, helium also has a lower opacity than hydrogen, decreasing the opacity. More importantly, these lower opacities cause the star to be more compact at a given mass, leading to higher effective temperatures, decreasing a star's lifetime. A complicating issue is that a variety of elements can be enhanced or depleted. In observational studies of Milky Way GCs, the effects of these enhancements affect the observed colours in different ways that are not easy to predict. For example, in a study of 47 Tuc ([Fe/H]=-0.76$\pm$0.02 \citep{CarrettaAbund}    ), helium enhancement of 0.04 and nitrogen  enhancement along with oxygen depletion has an effect on the F606W-F814W colour but not on the F336W-F814W colour in one scenario \citep{Milone2012}. However, the spread in the MS in the F606W-F814W colour is expected to be 0.02 mag, which is quite small. In another case for NGC 6397 ([Fe/H]=-2.0), helium and nitrogen enhancements of 0.02 do not show an offset in the F606W-F814W colour but an offset of 0.05 mag in the F336W-F814W colour, in direct contrast with the previous cluster \citep{NGC6397}.  

We can compare these results to previous analyses of multiple populations, as was done for NGC 1851 in \citet{Cummings2014} and \citet{Cummings2017}. In \citet{Cummings2014}, the authors perform a photometric analysis of NGC 1851 and discover a second, similar, and redder population on the MS as we have for Hodge 11 and NGC 2210. This is consistent with what was measured spectroscopically (\citet{Cummings2017} and references therein). The secondary population in NGC 1851 is C- and O-poor while N-rich and slightly enhanced in He. However, the lack of a distinct second population in the horizontal branch (HB) in NGC 2210 shown below indicates that NGC 2210 does not have a large difference in He abundance. Hodge 11 does show more evidence for multiple populations in the HB, but not as much as NGC 1851, indicating that there is some He enhancement but not as much as is seen in NGC 1851. \citet{Cummings2017}, however, illustrated that a split MS of these characteristics can easily result from CNO differences alone. Nonetheless, only a spectroscopic analysis will definitively show the abundance patterns of both of our GCs.

\section{Red Giant Branch Results}

The red giant branch (RGB) in these old GCs is characterized by low mass stars with a hydrogen burning shell around their helium core. Nearly all GCs have been shown to have split RGBs, especially when the `magic trio' of filters are used. This includes 47 Tuc \citep{Milone2012}, which has a colour spread of 0.1-0.2 dex in the F275-F336W colour. Their other colour combinations are not ideal to detect an RGB split caused by varying content of light elements. This means that in the colour combinations that we have access to in this work, the intrinsic spread in the RGB should be quite small, despite the presence of multiple populations.

\begin{figure*}
  \centering
  \begin{tabular}{cc}
    \includegraphics[width=\columnwidth]{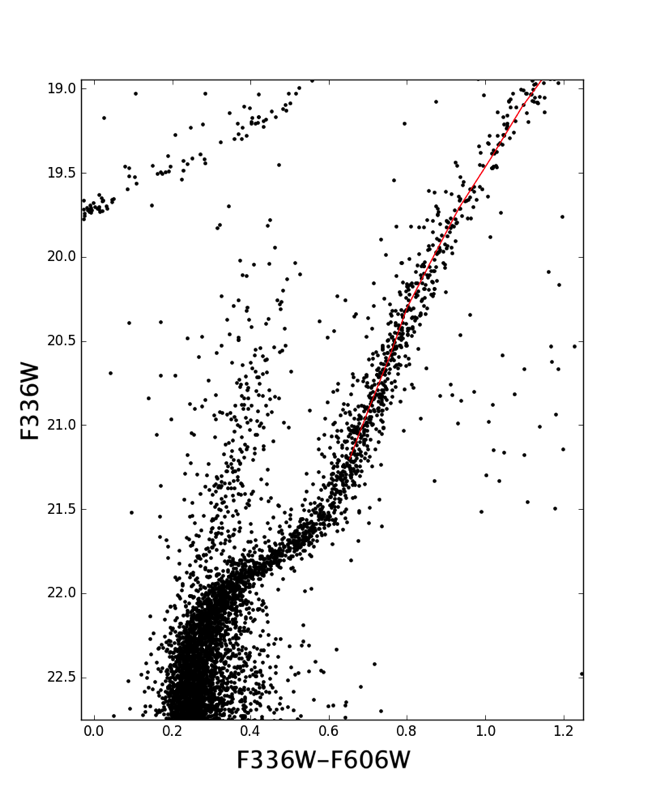} &
        \includegraphics[width=\columnwidth]{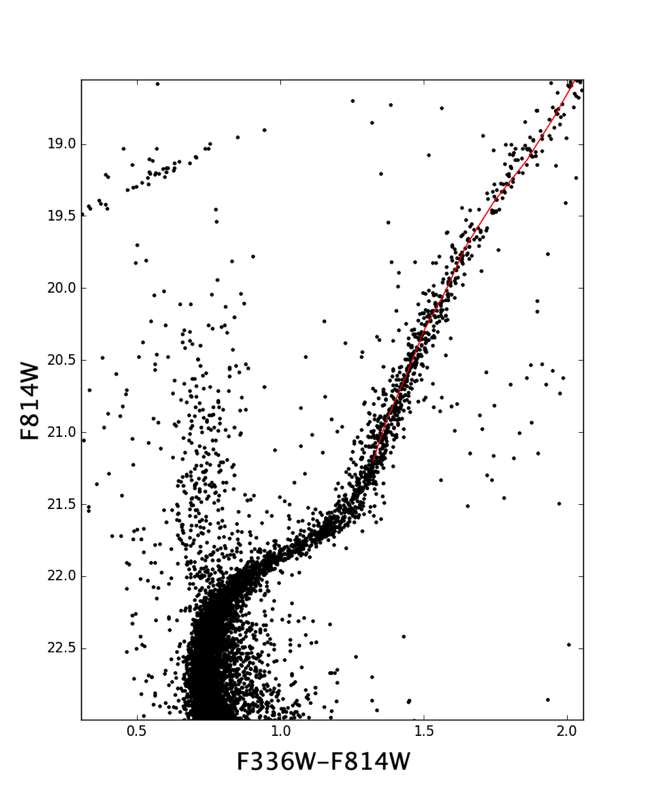}
 	 \\
 		                \small (a) &
                        \small (b)

  \end{tabular}
    \begin{tabular}{c}
  	\includegraphics[width=\columnwidth]{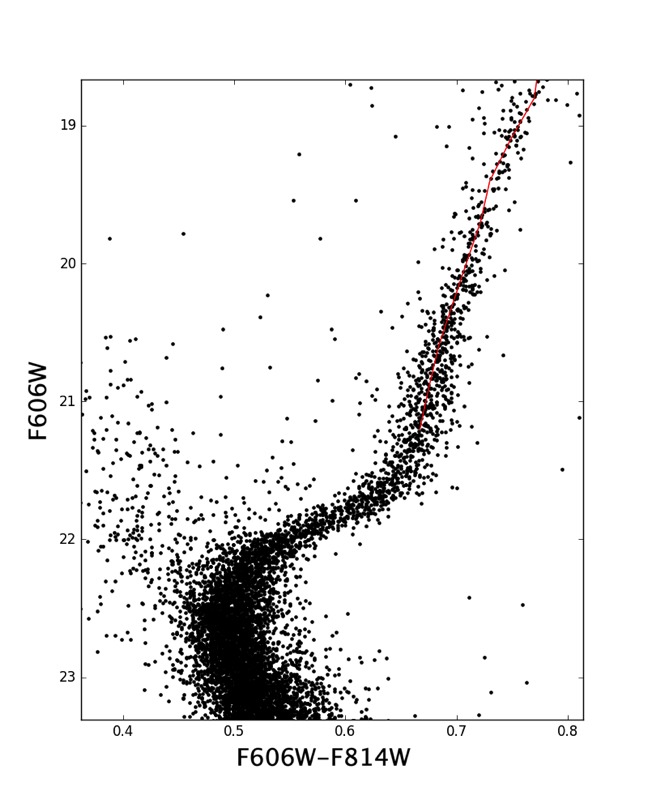}
  	\\
  	   \small (c) \\
  \end{tabular}
  \caption{These three panels show the colour-magnitude diagrams of the red giant branch (RGB) of Hodge 11 with the median ridge line shown in red. }
     \label{Hodge11RGB}
\end{figure*}

\begin{figure*}
  \begin{tabular}{cc}
    \includegraphics[width=\columnwidth]{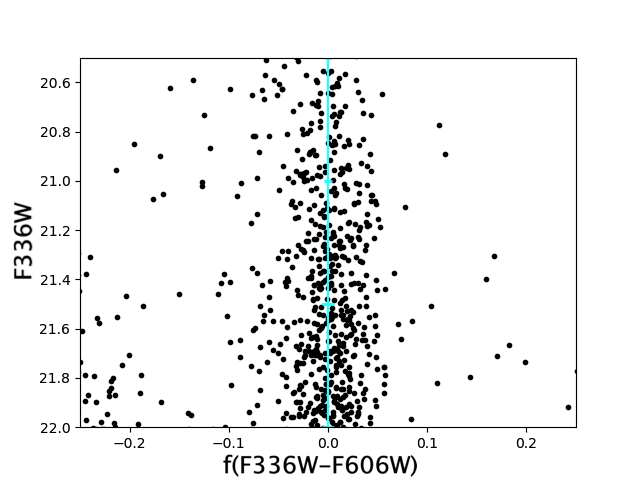} &
        \includegraphics[width=\columnwidth]{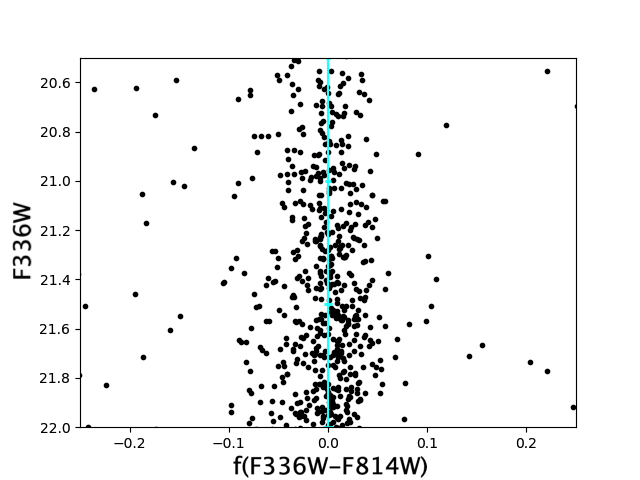} \\
	\small (a) & \small (b)
  \end{tabular}
\includegraphics[width=\columnwidth]{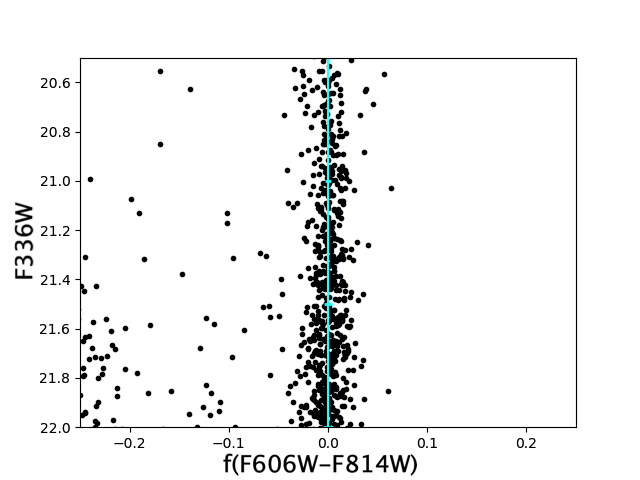} \\
\small (c)
  \begin{tabular}{c}

  \end{tabular}
\caption{These three panels show the straightened CMDs of Hodge 11 with respect to the median ridge line shown with characteristic errors in cyan. The error bars are present but are very small.}
\label{Hodge11RGBStraight}
\end{figure*}

\begin{figure*}
  \centering
  \begin{tabular}{cc}
    \includegraphics[width=\columnwidth]{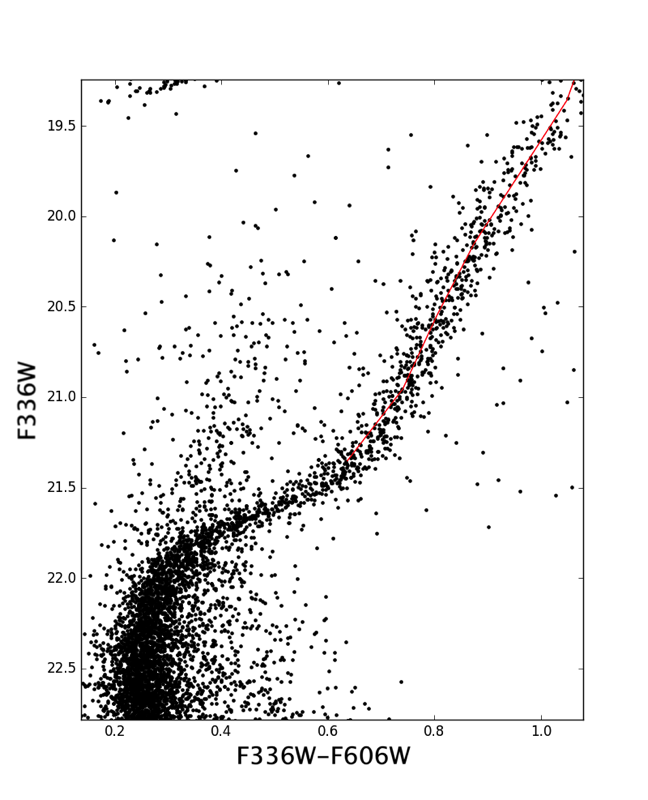} &
        \includegraphics[width=\columnwidth]{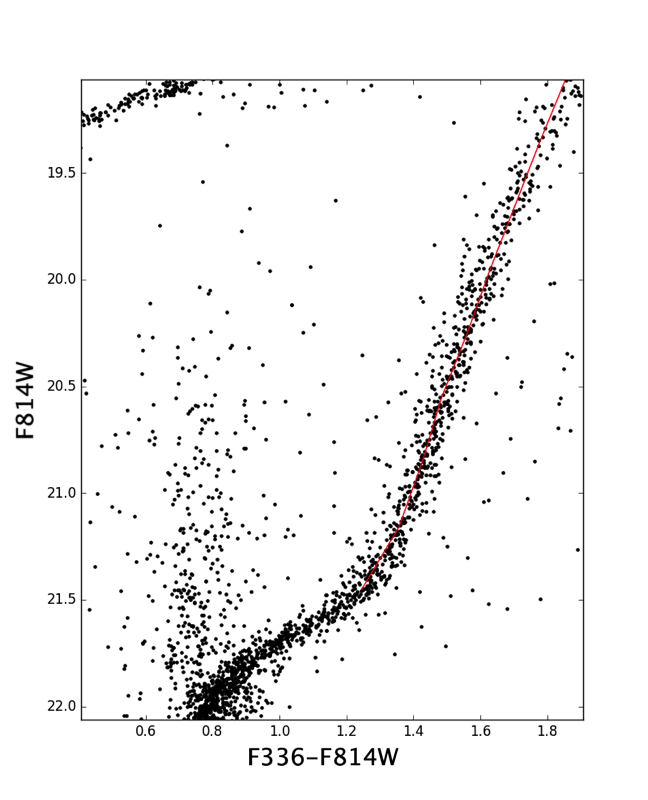}
 	 \\
 		                \small (a) &
                        \small (b)

  \end{tabular}
    \begin{tabular}{c}
  	\includegraphics[width=\columnwidth]{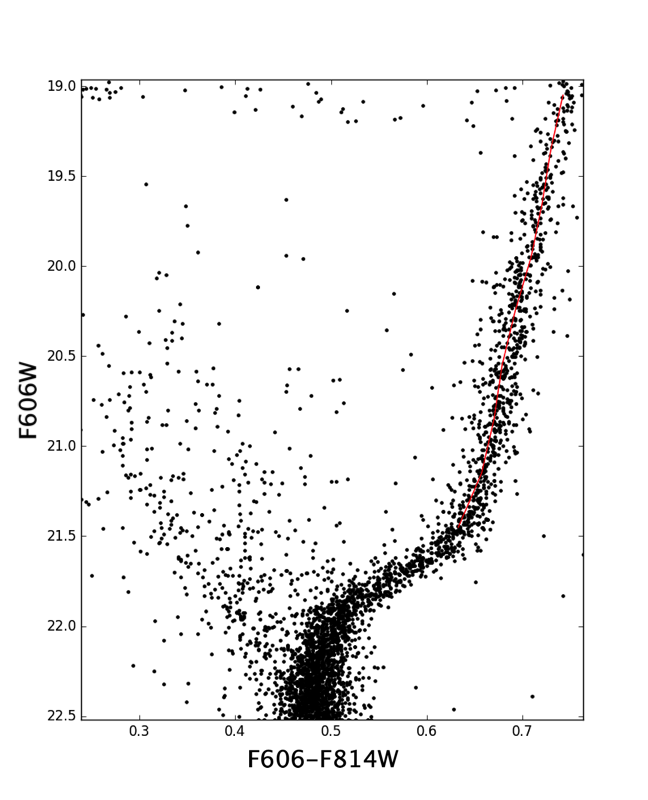}
  	\\
  	   \small (c) \\
  \end{tabular}
  \caption{These three panels show the colour-magnitude diagrams of the red giant branch (RGB) of NGC 2210 with the median ridge line shown in red. }
     \label{NGC2210RGB}
\end{figure*}

\begin{figure*}
  \begin{tabular}{cc}
    \includegraphics[width=\columnwidth]{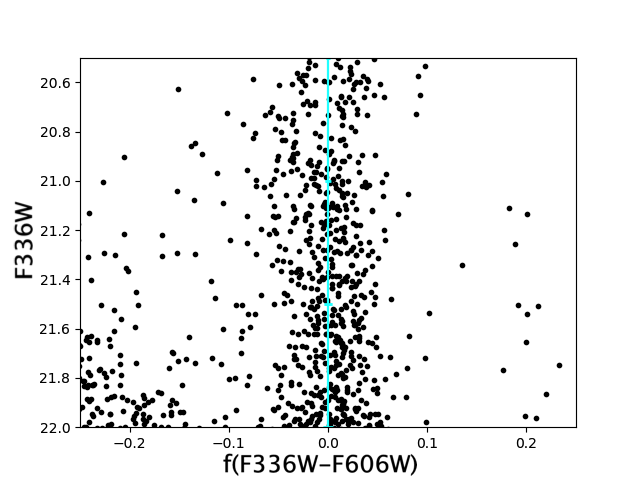} &
        \includegraphics[width=\columnwidth]{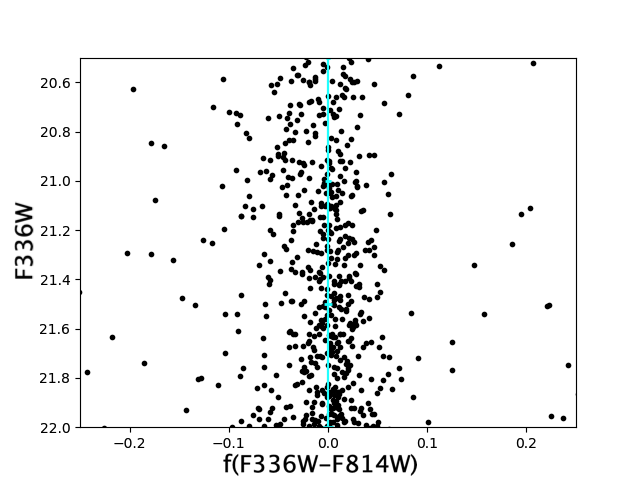} \\
	\small (a) & \small (b)
  \end{tabular}
  \begin{tabular}{c}
\includegraphics[width=\columnwidth]{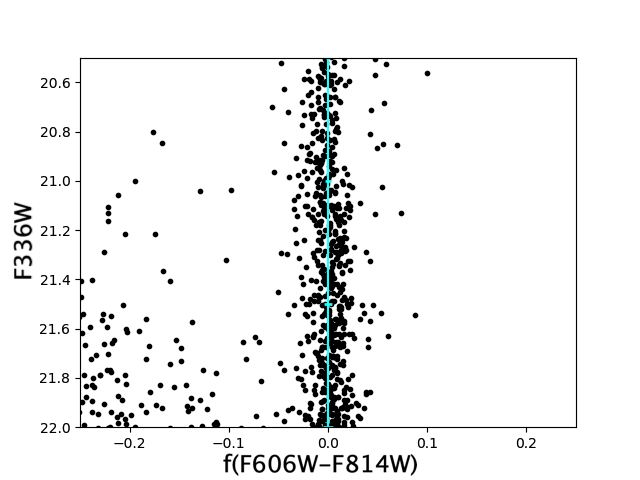} \\
\small (c)
  \end{tabular}
\caption{These three panels show the straightened CMDs of NGC 2210 with respect to the median ridge line shown with characteristic errors in cyan. The error bars are present but are very small.}
\label{NGC2210RGBStraight}
\end{figure*}

RGB branches with their median ridge line are shown in Figures \ref{Hodge11RGB} and \ref{NGC2210RGB} while the straightened CMDs are shown in \ref{Hodge11RGBStraight} and \ref{NGC2210RGBStraight}. The histogram analysis showed no evidence for a distinct second population in either of these clusters. However, it does seem that the F336W-F606W and F336W-F814W colours are wider than the F606W-F814 colour even after considering photometric errors. A more thorough discussion of the RGBs for our clusters is presented in \citet{Mackeyprep}.

\section{Horizontal Branch Results}

Both Hodge 11 and NGC 2210 have wide MS populations, as shown above. One may naively believe that this means they contain multiple populations with similar properties. However, by looking at each of their horizontal branches (HBs), it is clear that the clusters are quite different from each other. We perform isochrone fitting on each cluster using a wide array of isochrones, changing the age, metallicity, and helium abundance. Previous work \citep{Rachel} was used for starting properties, but the best fitting isochrones do vary from that work.  The final isochrones chosen for Hodge 11 and NGC 2210 are from the Dartmouth Stellar Evolution Program (DSEP) \citep{DSEP}. These isochrones provide a good fit to the shape of the main sequence and turn-off region. The distance modulus and reddening are then used to shift the ZAHB from the corresponding set of models. 

\begin{figure*}
  \centering
  \begin{tabular}{cc}
    \includegraphics[width=\columnwidth]{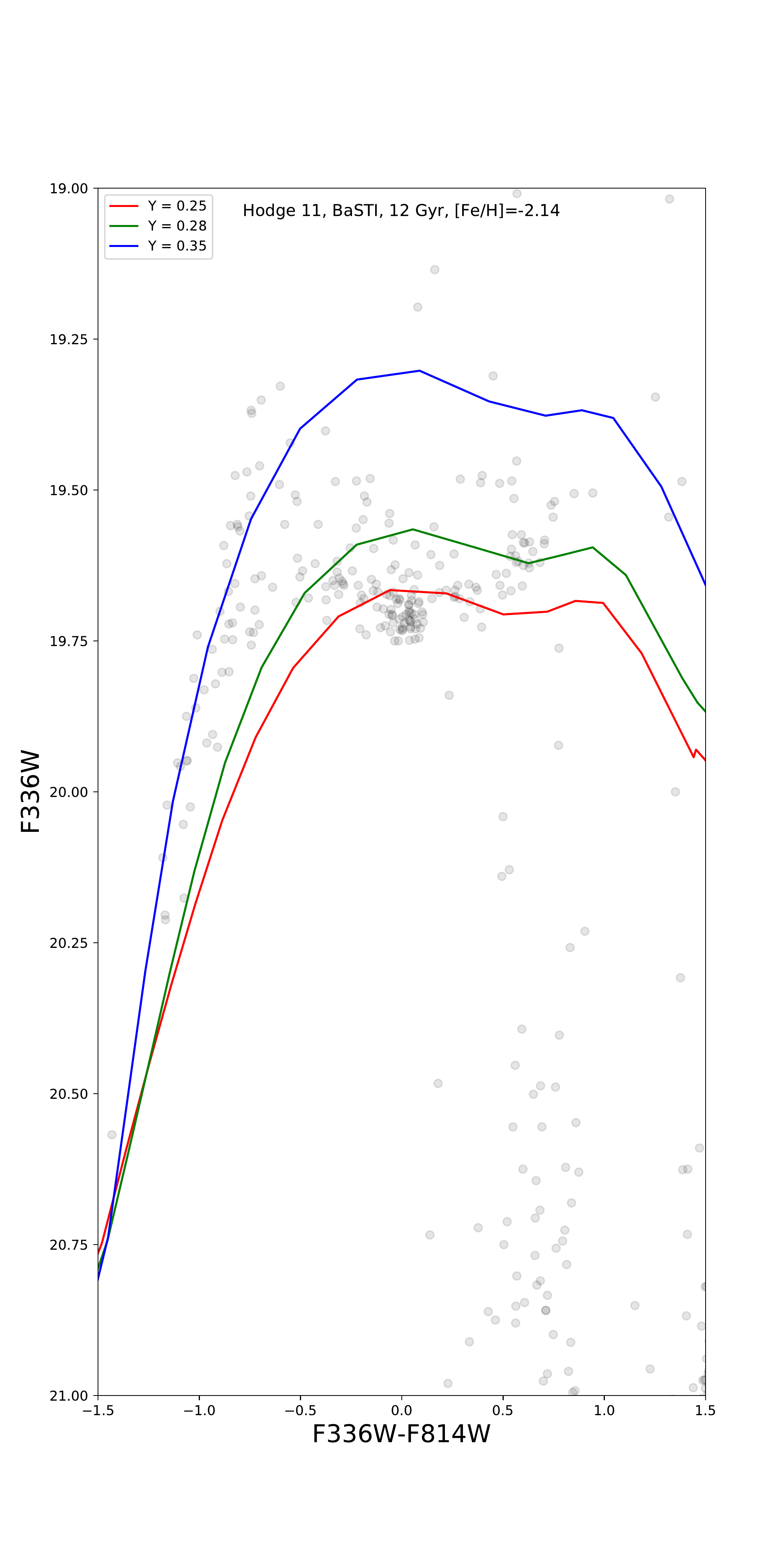} &
        \includegraphics[width=\columnwidth]{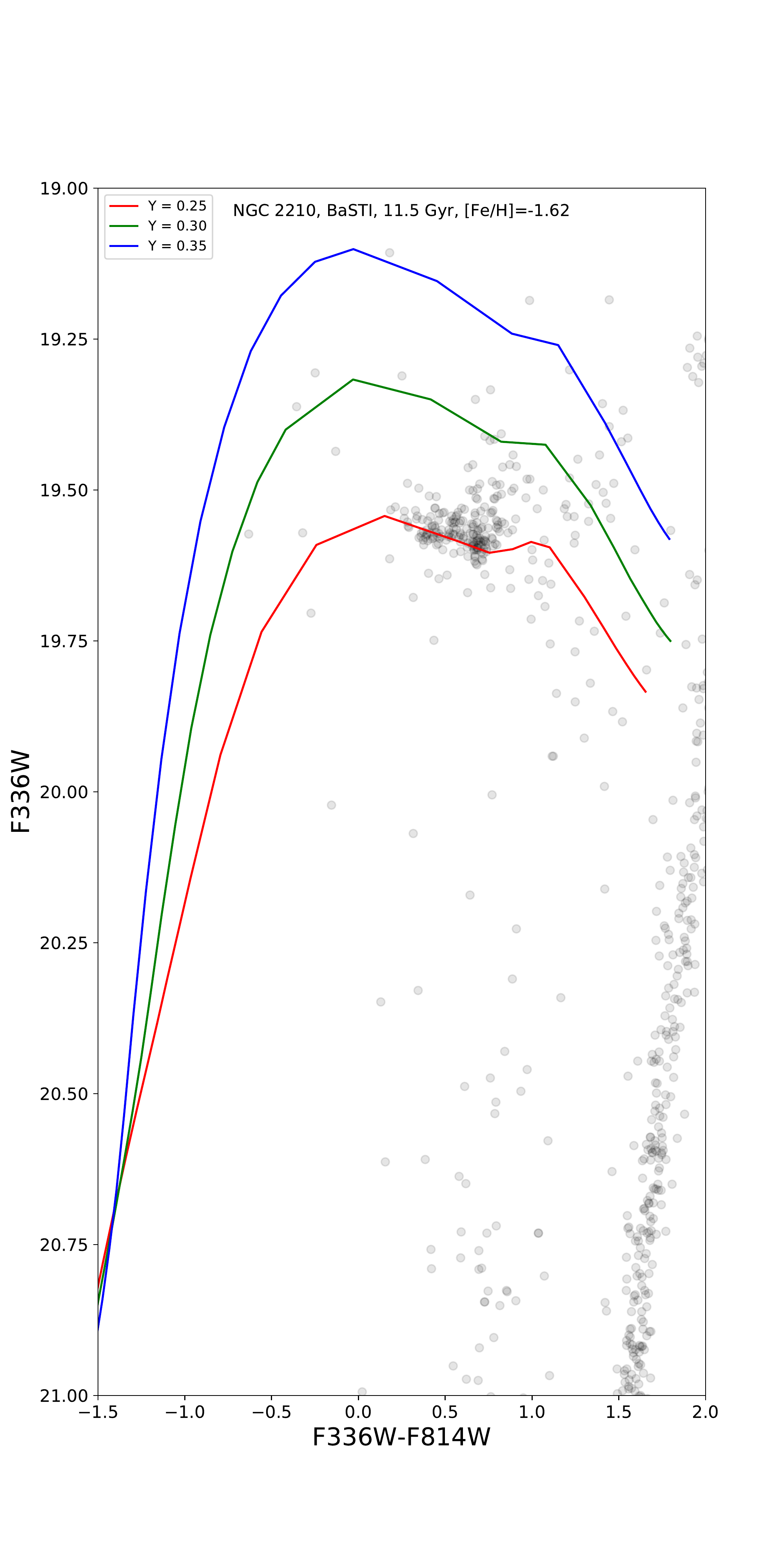} 
		\\
                \small (a) &
                        \small (b)
  \end{tabular}

  \caption{HB models of varying helium content from BaSTI using parameters found from an isochrone fit. For NGC 2210, the main clump of stars falls close to the primordial helium isochrone. The stars that lie above this clump do not seem to form a sequence and therefore are most likely not due to multiple populations. For Hodge 11, there are likely two different populations, one with a primordial helium abundance and the other enhanced by 0.1 dex.}
          \label{HB}
\end{figure*}

Helium has two contrasting effects on a HB star. At a given age, more helium-rich stars evolve faster, thus the mass evolving on the RGB belonging to a population enriched in helium is lower; this causes, on average, smaller masses of the envelope on the HB, which generally leads to lower luminosities. On the other hand, the higher the helium content, the higher the molecular weight of the H-burning shell. This renders the shell more efficient, thus leading to higher HB luminosities.

Theoretical HB models were constructed with the BaSTI code \citep{Basti}, as the Dartmouth code (used for the main sequence fitting) encountered numerical difficulties in evolving metal-rich HB models.  The BaSTI isochrones did not provide a good fit to the shape of the main sequence, making it difficult to use them to estimate the distance modulii and reddenings to the clusters. This mismatch between the main sequence and HB models could be of concern, but a detailed comparison of the Dartmouth and BaSTI models show that they are in general, in fairly good agreement with each other (e.g.\ \cite{Joyce15}). Since we are not performing detailed fits to the observed HB, this mismatch between the models used on the main sequence and the HB has a minimal impact on our qualitative discussion below.

We plot BaSTI \citep{Basti} HB models with varying He abundances to see where they lie in relation to the HBs. Varying model ages and metallicities did not reproduce the features seen in the HBs of the two clusters. The result for Hodge 11 is shown in Figure \ref{HB} Panel (a) and for NGC 2210 in Panel (b). The red ZAHB has a primordial helium abundance (0.25 dex) while the green is enhanced by 0.03 dex and the blue by 0.1 dex. There seems to be a 0.1 dex enhanced helium component for Hodge 11 while there is no enhanced population for NGC 2210. The most interesting aspect is that these two clusters' main sequences seem quite similar but their HBs are quite dissimilar. While Hodge 11 shows evidence of a  He enhanced population, NGC 2210's wide sequence does not seem to be He enhanced. This is similar to NGC 1851 \citep{Cummings2017} as mentioned earlier. NGC 1851 shows a stronger separation in the two populations on the HB, indicating that Hodge 11 has lesser He enhancement than NGC 1851, but it is still present. However, this is hard to reconcile with the redder MS of Hodge 11. If there was only He enhancement, the secondary population would actually be bluer than the primordial population. It is likely that CNO variations are overpowering the effect of the He difference in the F336W. However, further work, principally spectroscopic, needs to be performed to determine the precise abundance differences in these two GCs.

\section{Conclusion}
A detailed multiple population search in the old LMC GCs Hodge 11 and NGC 2210 found that both clusters exhibit a second MS population, with populations of 10\% and 18\% respectively after accounting for binary contamination. In addition, Hodge 11 also shows evidence for multiple populations in its HB. NGC 2210's HB does not show evidence for multiple populations, in spite of the similarity in the MS. This is the first photometric evidence that these ancient GCs in the LMC have multiple stellar populations. The RGB in the clusters also show some evidence of multiple populations. However, without spectroscopy, the chemical differences of the multiple populations cannot be known.  

\vspace{3mm}

This research has made use of NASA's Astrophysics Data System. Based on observations made with the NASA/ESA Hubble Space Telescope, obtained from the data archive at the Space Telescope Science Institute. STScI is operated by the Association of Universities for Research in Astronomy, Inc. under NASA contract NAS 5-26555. This work was supported in part by STScI through a grant HST-GO-14164. DG gratefully acknowledges support from the Chilean Centro de Excelencia en Astrof\' isica y Tecnolog\' ias Afines (CATA) BASAL grant AFB-170002. DG also acknowledges financial support from the Direcci\' on de Investigaci\' on y Desarrollo de la Universidad de La Serena through the Programa de Incentivo a la Investigaci\' on de Acad\' emicos (PIA-DIDULS). DM gratefully acknowledges support from an Australian Research Council (ARC) Future Fellowship (FT160100206). SV gratefully acknowledges the support provided by Fondecyt reg. n. 1170518.




\bibliographystyle{mnras}
\bibliography{LMC} 





\bsp	
\label{lastpage}
\end{document}